\documentclass[reprint,aps]{revtex4-1}
              \usepackage{dcolumn}
\usepackage{graphicx}

\usepackage{bm}
\usepackage{color}
\newcommand{\GeV}{\makebox{ GeV}}
\newcommand{\beq}{\begin{equation}}
\newcommand{\enq}{\end{equation}}
\newcommand{\beqa}{\begin{eqnarray}}
\newcommand{\beqast}{\begin{eqnarray*}}
\newcommand{\enqa}{\end{eqnarray}}
\newcommand{\enqast}{\end{eqnarray*}}
\newcommand{\nn}{\nonumber}

\newcommand{\alp}{\alpha_{\bf P}}
\newcommand{\req}[1]{(\ref{#1})}

\newcommand{\cT}{{\cal T}}

\newcommand{\al}{\alpha}
\newcommand{\be}{\beta}
\newcommand{\ga}{\gamma}
\newcommand{\de}{\delta}
\newcommand{\ep}{\epsilon}
\newcommand{\ze}{\zeta}

\newcommand{\la}{\lambda}

\newcommand{\si}{\sigma}

\newcommand{\ph}{\phi}

\newcommand{\om}{\omega}

\def\GeV{\nobreak\,\mbox{GeV}}

\begin{document}
\title{Diffractive Electromagnetic Processes from a Regge Point of View}
 %  \title{   Scale Dependent Pomeron Intercept in Electromagnetic Diffractive Processes  }
\author{H. G. Dosch}
\affiliation{Institut   f\"ur Theoretische Physik, Universit\"at Heidelberg \\
Philosophenweg 16, D-69120 Heidelberg, Germany     }
\author{E. Ferreira}
\affiliation{Instituto de F\'{\i}sica, Universidade Federal do Rio de Janeiro \\
C.P. 68528, Rio de Janeiro 21945-970, RJ, Brazil }

\begin{abstract}
The energy dependence of the cross sections for electromagnetic diffractive processes can be well  described by a single power, $W^\delta$. For $J/\psi$ photoproduction this holds in the range from 20 GeV to 2 TeV. This feature is most easily explained by a single pole in the angular momentum plane which  depends on the scale of the process, at least in a certain range of values of the momentum transfer. 
It is shown that this assumption allows a unified description of all electromagnetic elastic diffractive processes. We also discuss  an alternative model with an energy dependent dipole cross section, 
which is  compatible with the data up to 2 TeV and which shows an energy behaviour typical for a cut in the angular momentum plane.
\end{abstract}

\maketitle

\section{Introduction\label{intro}}
 Diffractive  processes involving virtual photons show a remarkable feature: the higher the photon virtuality $Q^2$,
 the faster is the increase of the cross sections with energy.
This feature is well understood in perturbative QCD, where the
evolution equations in $Q^2$
\cite{Gribov:1972ri,Altarelli:1977zs,Dokshitzer:1977sg} predict
such  a behaviour; the rising rate of increase in energy
can be traced back to the increase of the gluon density with
higher resolution.

This
 specific feature of the energy dependence,  however, is less  easily explained  in Regge theory~\cite{Col77}. The underlying  core concept of this theory is the Sommerfeld-Watson transform~\cite{Wat19,Som49}. The sum over the partial waves of a scattering amplitude in the $t$ channel is replaced by a contour integral in the angular momentum plane. The high energy behaviour of a process in the $s$ channel is determined by the position of the singularity in the complex angular momentum plane farthest to the right. If the singularity is a pole at position $\ell=\al$ the high energy behaviour of the amplitude is $\cT \sim (W/W_0)^{2 \al}$.  If it is a branch cut at $\ell= \al_C$ the power behaviour, up to logarithmic terms, is ultimately driven to  $\cT
\sim (W/W_0)^{2 \al_c}$, but the explicit form depends crucially on the
behaviour of the discontinuity across the cut.

 Usually  the positions of the singularities in the complex plane are assumed to be independent of the specific  process.
 In purely  hadronic diffractive processes the energy dependence of the scattering amplitude is supposed to be determined by the  position $\alp(t)$  of  a specific singularity, the ``Pomeron trajectory"  which depends on  the squared momentum transfer $t$.  Based on a large amount  of hadronic diffractive  data, Donnachie and Landshoff \cite{Donnachie:1992ny} proposed a general description with  $\alp(0) \approx 1.09$ and a
slope  ${d \alp}/{dt} =\alp ' = 0.25 \GeV ^{-2}$.

 On the other hand,   by summing up leading-log terms in perturbative  QCD,  a Pomeron with a larger value  $\al_P(0)$ was found (BFKL-Pomeron)~\cite{Fadin:1975cb,Kuraev:1977fs,Balitsky:1978ic,Ciafaloni:1998gs,Fadin:1998py}. 
%                    ,Brodsky:1998kn}.
 Donnachie and Landshoff \cite{Donnachie:1998gm} extended the Pomeron concept and assumed that electromagnetic diffractive  processes are determined by two Pomerons, a soft (hypercritical) one with $\alp(0)=1.09$ and a hard one with   a value of  1.42. This idea has been applied in many electroproduction processes, where the couplings to the two Pomerons were essentially determined  by the size of the scattered objects and hence in a given model the energy dependence was universally fixed by a superposition of the two Pomeron contributions. In this way a comprehensive description of  proton structure functions, vector meson production
and $\gamma^*$-$\gamma^*$ scattering could be achieved  in the full
energy  range accessible at HERA
\cite{Dosch:1994ym,Dosch:1997nw,Kulzinger:1998hw,Donnachie:1999kp,
Donnachie:2000px,Donnachie:2001wt,Dosch:2002ig,Dosch:2006kz,Baltar:2009vp}.

Recent experiments  of $J/\psi$ photoproduction at LHC at energies
up to the TeV region \cite{LHCb(14),Alice(14)}
 have shown, however, that a single power,
corresponding to $\alp(0)= 1.17$ describes very well the energy
dependence in the range from 20 GeV to 2 TeV.  
This behaviour, although with larger 
experimental uncertainties, has also been found in $\Upsilon$ photoproduction
 \cite{LHCb(15)}. 
These results are hardly compatible with the two Pomeron picture and rather support 
the concept of  a single singularity in the angular momentum plane
determining the high energy behaviour.

Our paper is organized as follows. In Sect. \ref{scale1} we
discuss the possibility of scale-dependent singularities in the
complex angular momentum plane and
define scales which allow to relate the energy dependence of
vector meson production cross sections to the $x$-dependence of the 
proton structure function. In Sect. \ref{edip} we consider a model for
$\ga^*\,p$ scattering and diffractive vector meson production
where the energy dependence is due to a specific energy dependence
of the dipole cross section. In Sect. \ref{data1} we   compare the results of both models with
experiment. Finally in Sect. \ref{final} we compare 
the two approaches and discuss the implications on the Regge picture.

This paper is an extension of our earlier preprint {\it Scale-Dependent Pomeron Intercept in Electromagnetic Diffractive Processes}, arXiv:1503.06649 [hep-ph], and replaces it.

 \section{Scale-dependent Regge singularities \label{scale1}}

\subsection{General considerations \label{general}}
\begin{figure}
\begin{center}
\includegraphics*[width=4cm]{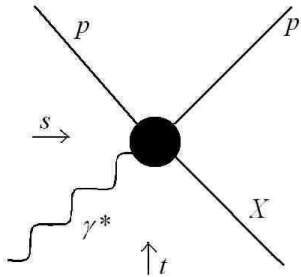}
\end{center}
\caption{ \label{reaction}Virtual photon scattering ($X= \ga^*$)
or diffractive vector meson electroproduction ($X$ denotes a
vector meson) off protons.}
\end{figure}

We  sketch  essential features of the Sommerfeld-Watson transform,
neglecting  details  needed to include spin and
signature effects. We consider the reaction visualized in  Fig.
\ref{reaction}, involving a virtual photon $\ga^*$, two protons $p$
and a particle $X$ with the same quantum numbers as  $\ga^*$; the
latter  can be  either be  a virtual photon or a vector meson.

In an obvious notation  we denote by $p_{\ga^*},\, p_p,\,p_X$ the
momentum of the virtual photon, the incoming proton, and the
particle $X$, respectively;  $Q^2=-p_{\ga^*}^2$ is the photon
virtuality, $m_p$ the proton mass and $m_X^2 = p_X^2$ . The
squared CM energy and momentum transfer in the $s$ channel are  $s
=(p_{\ga^*} + p_p)^2, \; t=(p_X- p_{\ga^*})^2$; $\cT(Q^2,s,t)$ is
the scattering amplitude involving these particles. In the $s$
channel the amplitude $\cT(Q^2,s,t)$ describes the process $\ga^* \, p \to X \,
p$,  in the $t$ channel it describes the process $\ga^* \,\bar X \to p \bar p$.

 The partial wave representation of this amplitude in the $t$ channel is given by
\beq \label{suml} \cT(Q^2,s,t) = \sum_\ell (2 \ell+1) ~ 
\cT_\ell(Q^2,t) ~ P_\ell(z_t) ~ ,  \enq 
where $z_t$ is the cosine of  the
CM scattering angle in this  channel, \beq z_t= \frac{\sqrt{t}\;
(2 s +t -2m_p^2-m_X^2 + Q^2)}{\sqrt{\big(t^2 - 2 t(m_X^2 -Q^2)
+(m_X^2+Q^2)\big)\, \big(t-4 m_p^2\big)}} ~.  \enq
By the Sommerfeld-Watson transformation the sum in Eq. \req{suml} is
expressed as the contour integral 
\beq \label{intl} \cT(Q^2,s,t) =
\int_C d\ell \,  (2 \ell+1) ~ \cT_\ell(Q^2,t)~ 
\frac{P_\ell(z_t)}{\sin \pi \ell} ~ . \enq 
Since for large $s$ and
fixed $t$ and $Q^2$ the quantities $z_t$  and  $P_\ell(z)$ behave like 
  $z_t \sim s$  and $P_\ell(z) \sim
z^\ell$, the high energy behaviour of $\cT$  is determined by the
position of the singularity in the $\ell$ plane with the largest
value of $\Re \ell$.
If this singularity is a pole at position $\ell = \alp(Q^2,t)$
then the high energy behaviour is \beq \cT(Q^2,s,t) \sim s^{
\alp(Q^2,t)} ~ , \enq
and in electromagnetic diffractive processes its position has to depend on 
the photon virtuality $Q^2$ in order to
be compatible with the data as described above.
 Therefore the assumption of a universal position of the singularities in 
the complex angular momentum plane for all virtualities has to be abandoned 
in this case.  This does not preclude the possibility that in the $t$ channel, 
$t > 0$,  there might exist  hadrons corresponding to the poles in the 
angular momentum plane,and also, for instance, glueballs. We  discuss such scenario in 
Sect. \ref{final}.
Additional motivation to suggest a scale-dependent Regge
trajectories in the scattering domain came from holographic models
for diffractive reactions \cite{Brower:2006ea,Hatta:2007he}.

Without the assumption of universal singularities in the complex
angular momentum plane for electromagnetic processes,  Regge
theory looses much of its predictive power in this field. One may
venture, however, to postulate that the position of the
singularity depends only on the scale of the specific reaction,
but not on the process itself. The fact that  Regge poles seem to
be universal for all processes where only the hadronic scale is
involved, including real-photon nucleon scattering, supports such
an assumption. In order to test the hypothesis of universal 
scale-dependent Regge singularities in non-purely hadronic processes, 
we have to find a relevant scale and
a way  to match it for different reactions, like deep
inelastic scattering and diffractive vector meson production.
Generally, there is also the possibility that high energy elastic
$\ga^*$\,p scattering and diffractive vector meson production are
not determined by a pole but by a branch cut in the complex angular
momentum plane. If the branching point of such a singularity is at
$\ell = \al_C$ the high energy behaviour up to logarithmic
terms is eventually given by $s^{\al_c}$ but  how fast this behaviour
is approached depends strongly on the discontinuity at the cut.
In Sect. \ref{edip} we  explore a model which yields 
 the high energy behaviour
of diffractive electromagnetic processes  determined by a cut.
The position of this cut could well be universal, but the
discontinuity would be scale-dependent.

\subsection{Defining  scales  for different processes  \label{scale_sec}}

The structure function $F_2(Q^2,x) $ of deep inelastic scattering
is the best investigated diffractive quantity. It is related to the
$\ga^*\,p$ total cross section by
  \begin{equation}
F_2(x,Q^2) = \frac{Q^2}{4 \pi^2 \, \al} \si_{\rm tot}^{\ga^*p}  ~,
  \end{equation}
  with
\begin{equation}
   x=\frac{Q^2}{W^2 + Q^2 -m_p^2} ~ .
  \end{equation}
For energies in the HERA range  and $Q^2 > 1 \mbox{ GeV}^2$ the structure functions ~\cite{Adloff:2001rw}  can be
fitted by a single power \cite{Radescu:2013mka}
 $F_2(x,Q^2)= c ~ x^{-\la(Q^2)}$  with 
$\la(Q^2) = 0.0481 \log\left[ Q^2/0.0853 \; \rm{GeV}^2  \right]$ .

 Due to the optical theorem  the  $\ga^*\, p$ cross section is proportional
to  the $\ga^*\, p$ forward scattering amplitude
$\cT_{\ga^*p}(Q^2,W^2,0)/W^2$ and the $x$-dependence mentioned  above
leads to  the energy dependence $\cT_{\ga^*p}(Q^2,W^2,0) \sim
(W^2)^{1+\la(Q^2)}$. This behaviour corresponds to a pole in the
angular momentum plane at position $\alp(0)= {1+\la(Q^2)}$. Thus the
"effective power" can be interpreted as the position of a $Q^2$-
dependent pole (Pomeron pole) in the angular momentum plane.

We shall use the modification
\begin{equation}\label{laga} \lambda(Q^2)  = \al_{\rm \bf P}(0)-1= 0.0481 \log\left[\frac{Q^2+0.554}{0.0853}\right] ~ ,
\end{equation}
which is adjusted to give the intercept 1.09 at hadronic scales,
that is at $Q^2=0$.

In a space-time picture  the virtuality $Q^2$ of the
virtual photon $\ga^*$ is related to the size of its hadronic
structure.
 The planar quark density of the hadronic  light-front wave function of a virtual photon 
 can be derived from perturbation theory. For  photons with transverse polarization we obtain
 \begin{eqnarray}\label{rhotr}
&& \rho_{\ga^*\ga^*;\pm1}(Q^2,u,b_\perp) = \hat e_f^2\frac{6 \al}{4 \pi^2}\,b_\perp \\
&&  \left[ (Q^2 u(1\!-\! u) + m_f^2) (u^2 \! +\! (1\! -\! u)^2
)\, K^2_1(\ep b_\perp)+\! m_f^2\,K^2_0( \ep b_\perp)\right],
\nonumber
\end{eqnarray}
and for longitudinal polarization
\begin{eqnarray} \label{rhol}
&& \rho_{\ga^*\ga^*;0}(Q^2,u,b_\perp)    \\ \nonumber && =\hat
e_f^2\frac{12 \al}{4 \pi^2}  \, b_\perp ~  Q^2 \,u^2(1-u)^2 \,
K^2_0(\ep b_\perp)  ~ ,
\end{eqnarray}
where
\begin{equation}
 \ep = \sqrt{Q^2 {u(1-u)} + {m_f^2}\,} ~ ;
\end{equation}
 $u$ is the longitudinal momentum fraction of the quark, $b_\perp$  the  transverse 
separation between the quark and the antiquark,  $m_f$   the  mass of the quarks, and 
$\hat e_f$ is the effective charge.

It is intuitive to assume that the ``size" of the virtual photon
sets the relevant scale. Since the planar density $
\rho_{\ga^*\ga^*;\pm1}$ in  Eq. \req{rhotr}  is not normalizable,  we
cannot define a mean square radius in the usual way. We then define  as
scale $\bar b$  the value where the expression 
\beq \label{Y}
 Y(Q^2,b_\perp)=b_\perp^2 \,  \int_0^1 du \,  \rho_{\ga^*\ga^*; \rm pol}(Q^2,u,b_\perp)
\enq 
is maximal,  
\beq \label{Ymax} \bar b = \max_{b_\perp}
\,Y(Q^2,b_\perp) ~ .\enq
 
For vector-meson electroproduction  we take analogously  as scale
the maximal value for the corresponding expression of  the
overlap between the photon and meson wave function. In the 
transverse case the
planar overlap  density is given by 
\begin{eqnarray}\label{vmtr}
&& \rho_{\ga^*,V;\pm1}(Q^2,u,b_\perp) = \hat e_V \frac{\sqrt{6 \al}}{2 \pi} \,b_\perp\,\ph_{\omega}(u,b_\perp)   \\
&& \left[ 4 \ep\, b_\perp  \, \om^2 (u^2 +(1\! - \! u)^2 )\, K_1(
\ep b_\perp) \! +\! m_f^2\,K_0(\ep b_\perp) \right],
  \nonumber
\end{eqnarray} 
 and for longitudinal polarization
\begin{eqnarray} \label{vml}
&& \rho_{\ga^*,V;0}(Q^2,u,b_\perp ) \\
&& = 16 \hat e_V \frac{\sqrt{3 \al}}{2 \pi}\, \,b_\perp  \om \, Q
\, u^2 (1-u)^2\,K_0(\ep b_\perp) ~  \ph_{\omega}(u,b_\perp) ~ ,
\nonumber
\end{eqnarray}
 with   mass $m_f$ for  the quarks constituting the vector meson, 
and  $\omega$ accounting for the wave function width. 

For the meson wave functions $\ph_{\omega}(u,b_\perp)$,  we use  the the
Brodsky-Lepage (BL) ~\cite{lep80}  form 
  \beqa\label{BL} \ph_{\omega}(u,b_\perp) &=&
\frac{N}{\sqrt{4 \pi}} \times \\ \nn
&&\hspace{-1cm}\exp\left[-\frac{m_f^2(u-1/2)^2}{2 u (1-u)
\om^2}\right]
 \exp[-2 \om^2  u (1-u) b_\perp^2] ~. 
\enqa
 For convenience, the values of N and $\omega$ in the BL wave 
function  (\ref{BL}) determined by the electronic decay widths 
 \cite{Dosch:2006kz,Baltar:2009vp}  are given    in Table
\ref{WFparam} in Appendix 1.

 The planar densities $\rho(Q^2,u,b_\perp)$ depend on  the quark masses.
For diffractive production of heavy vector mesons we use the
$\overline{MS}$ masses \cite{Agashe:2014kda}: $m_c=1.28$ GeV and
$m_b=4.18$ GeV. For $Q^2 =0$  the overlap  diverges
logarithmically with vanishing quark mass and therefore special
constituent mass values have to be assumed. In order to reduce
model dependence,
 we have for light meson production determined the scale only for 
$Q^2 \geq 1$, where the dependence
on quark masses is weak and the current quark masses,
 $m_u \approx m_d \approx  0$, $m_s=0.1 $ GeV can be safely chosen.
For hadronic processes involving light quarks, the scale at
$Q^2=0$  is fixed by the confinement scale and therefore we have
there  the purely hadronic Pomeron intercept $\al_{\bf P}(0) \approx
1.09$.

\begin{figure}
\begin{center}
\includegraphics*[width=8cm]{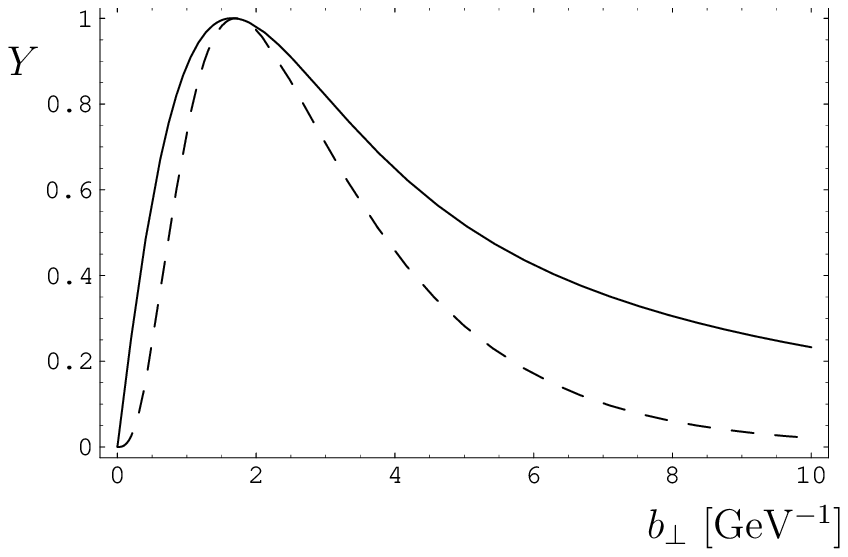}
\end{center}
\caption{ \label{overlaps} The function $Y(Q^2,b_\perp)$, Eq.
(\ref{Y}), as function of $b_\perp$, for $\rho_{\ga^* ,\ga^*;1}$
at $Q^2=2$ \GeV$^2$ (solid curve) and for   $\rho_{\ga^*\,
\rho;1}$ at $Q^2= 34$ \GeV$^2$ (dashed curve). For exemplification 
purpose, the choice of parameters is made so that the peaks coincide. }
\end{figure}

Typical forms of the function  $ Y(Q^2,b)$, Eq.\req{Y},  for
transversely  polarized photons and $\rho$ mesons , normalized to 1 at the maximum
are shown in Fig.
\ref{overlaps}. In the example, the  $Q^2$ values are chosen so 
that the  peaks at $\bar b$ coincide.

\begin{figure}
\begin{center}
\includegraphics*[width=8cm]{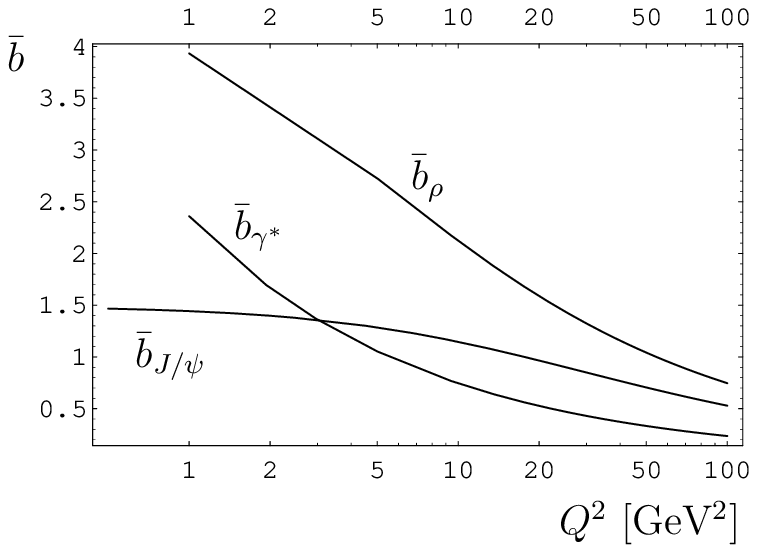}
\end{center}
 \caption{ \label{scales_fig} Plots of   the scales
  $\bar b(Q^2)$
  that convert $Q^2$ into the scale variables that are
    used to  describe  different diffractive
   processes of  $\rho$ and $J/\psi$  vector meson production
and  photon scattering.
 }
 \end{figure}

 In  Fig. \ref{scales_fig} we show the scales  $\bar b (Q^2)$, obtained as the $b_\perp$ value where the function (\ref{Y}) is maximal
for  $\rho$ and $J/\psi$  vector meson production and for  photon
scattering as function of $Q^2$.

 In order to relate the position of the pomeron pole with the scale $\bar b$ one
inverts the scale function $\bar b_{\ga}(Q^2)$, obtained  for
$\gamma^*\, p$ scattering; this yields the inverse function
$\overline{Q^2}_{\ga}\big(\bar b \big)$.  We can then  calculate
the value of the pomeron pole position for  vector meson
production as function of $Q^2$ from Eq. \req{laga} by inserting for
$Q^2$ the value $\overline{Q^2}(\bar b_{VM})$, where $\bar b_{VM}$
is the  value $\bar b_{\rm VM}(Q^2)$ obtained for the vector meson
at photon virtuality $Q^2$.  We thus obtain  for the production of
the vector meson  VM the relation
\begin{equation}\label{laga2}
  \al_{\rm \bf P}(0)-1= 0.0481 \log\left[\frac{\overline{Q^2}_\ga\big(\bar b_{\rm VM}(Q^2)\big)+0.554}{0.0853}\right] ~ .
\end{equation}

The results of the numerical analysis of the scales -- the maxima
$\bar b$ of the function (\ref{Y}) --  show that for each process
$\ga^* \, p \to ({\it fs}) \, p$ with  final state ({\it fs})
and given polarization "pol", the average scale $\bar b (Q^2)$ can
be very well fitted by  a function of the simple form 
\beq
\label{fit}
 \bar b_{\it fs,\rm pol}(Q^2) = \frac{A_{\it fs,\rm pol}}{\sqrt{Q^2 + B_{\it fs,\rm pol}}} ~.
\enq
The values for the constants $A$ and $B$ are collected in Table
\ref{res} in Appendix 1.

From this fit and Eq.\req{laga2} we can obtain the Pomeron pole
position  from $\ga^*$ scattering as a function of the scale $\bar
b $.  We then have
\begin{eqnarray}\label{laga3-b}
 &&   \al_{\rm \bf P}(0)-1  \\
&& = 0.0481 \log\left[\frac{A_{\gamma^*,\rm{pol}}^2/ 
{\bar b}^2 -B_{\gamma^*,\rm{pol}}+0.554}{0.0853}\right] ~ , \nonumber 
  \end{eqnarray}   
where $A_{\gamma^*,\rm{pol}}$  is the  coefficient in Eq.\req{fit}
for $\ga^*\,p$ scattering, with the label pol indicating
transverse (T), longitudinal (L)  or total (tot) cross sections
(row  $\ga^* $ in Table \ref{res}).

From this equation we obtain the intercept for vector meson
production as function of $Q^2$ by expressing the scale $\bar b$
through Eq.\req{fit} for the specified meson  
\begin{eqnarray}  \label{meq}
\delta(Q^2)&=& 4 ( \al_{\rm \bf P}(0)-1) = 0.472+ 0.1924 \times\\
&&\hspace{-1cm}  \log\left[\frac{A_{\gamma^*,\rm{pol}}^2}{A_{\rm
VM,pol}^2} (Q^2 -B_{\gamma^*,\rm{pol}}+B_{\rm VM,pol})
+0.554\right] ~.    \nonumber
 \end{eqnarray}
The functions $\de(Q^2)$ for the different  processes of
vector meson production with polarization ``pol'' are listed in Table
\ref{del} in   Appendix 1.

 The pole position  at $t=0$ determines the energy behaviour of the forward scattering
amplitude (and, therefore,  also of the total  $\ga^* \,p$ cross
section). For integrated elastic production cross sections we 
also have to take into account the $t$ dependence of the Regge
singularity, which generally leads to a shrinkage of the
diffraction peak as the energy increases. For unpolarized elastic diffractive vector meson
production, $\ga^*\, p \to (V\!M) \, p $, the differential elastic
cross section in the Regge model  is given by \beq
\frac{d\si}{dt}=\left(\frac{s}{s_0}\right)^{2 [\al_{\bf P}(t)-1]}
\be^2(t) ~ . \enq For fixed $W$ and $Q^2$ the $t$ dependence is
well approximated by an exponential.  We  assume the residue
$\be(t) = \be_0 e^{D \,t/2}$ and
 $\al_{\bf P}(t) = \al_{\bf P}(0) + \al'_{\bf P}\, t $.  We then obtain for the
integrated   cross section
\begin{eqnarray}
&& \si_{\rm int} =\int_{-\infty} ^{0}  dt\,\frac{d\si}{dt} \\
&&  = \frac{\be_0^2}{D + 2\,\al'_{\bf P}\log(s/s_0)}\,
\left(\frac{s}{s_0}\right)^{2[\al_{\bf P}(0)-1]}\left(1 +
O(s^{-2})\right) ~.  \nonumber
\end{eqnarray}
The slopes observed in $d\sigma/dt$ in vector meson
electroproduction  \cite{Dosch:2006kz} are in the range from  5 to 10
GeV$^{-2}$. With $2 \, \al'_{\bf P}/D \ll 1$,  the energy
dependence of the total cross section can be approximated by \beq
\label{shrink2} \si_{\rm int} \approx \frac{\be_0^2}{D}
\left(\frac{s}{s_0}\right)^{2[\al_{\bf P}(0) -  \al'_{\bf P}/D
-1]} ~ . \enq

Although  the present  data on $\al'_{\bf P}$ do not allow firm
conclusions~\cite{H1(10)}, it is certain that the effective powers
$\delta_{\rm int}$  that fits experiments should be smaller  than
the value $4 [\al_{\bf P}(0) -1] $ obtained from the structure
function. Inspired by  a simplified  model discussed in
~\cite{Brower:2006ea} we make the ansatz that the slope of the
Pomeron singularity  decreases with decreasing scale
    \beq
\label{shrink1a} \al'_{\rm P} = \al' \frac{\bar b^2}{\bar b_{\rm
conf}^2} ~ , \enq
     where $\bar b_{\rm conf}$ is the
scale set by confinement, at which $\al'_{\rm P}\approx 0.25 ~
$GeV$^{-2}$.
 Choosing   realistic values $\bar b_{\rm conf}= 5 $  GeV$^{-1}$,
$D=5$ GeV$^{-2} $,   we obtain a shrinkage  correction
  \beq \label{shrink3} \frac{\al'_{\rm P}}{B} =
0.002 \,\bar b^2 = 0.002 \, \frac{A_{\rm VM,pol}^2 }{Q^2 +B_{\rm
VM,pol}} ~, \enq and  for the power $\delta_{\rm int}$, applicable
to integrated elastic diffractive cross sections we then have
  \beqa
\label{delta-int}
\de_{\rm int}(Q^2) &=& \de(Q^2) - 4 {\al'}_{\rm P}/B \\
&=&\de(Q^2)  -0.008 \; \frac{A_{\rm{VM,pol}}^2 }{Q^2 +B_{\rm{
VM,pol}}} ~ ; \nn \enqa
    the functions $\de(Q^2)$ for the
different processes  are defined in Eq. (\ref{meq})  and displayed
explicitly in Table \ref{del}. The shrinkage corrections 
Eq. (\ref{shrink3})  are mostly very small, except for photoproduction of $\rho$
  mesons, where they reduce the power $\de$ from the soft 
pomeron value 0.36 to the observed value of about 0.19. 

It must  be noted that the absolute value of the scale $\bar b$
plays no role. Only the relation between the scale for $\ga^*\,p$
scattering and the scale for vector-meson production, which leads
to the relation \req{meq} is of phenomenological relevance. There
might be different choices of the scale  leading   to similar
results.

\begin{figure}
\begin{center}
\vspace{1cm}
\includegraphics*[bb= 17 150 525 654, width=8cm]{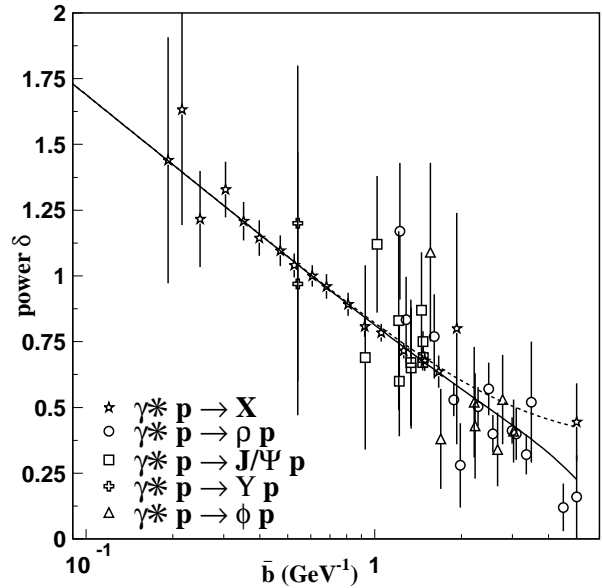}
\end{center}
\caption{ \label{all} Experimental values of $\delta = 4(\al_{\bf
P}(0)-1) $ vs. the scale $\bar b $ for different processes.  The
dashed line represents  the interpolation formula \req{laga}. The
solid curve takes into account the effects of the shrinking in
Eq.\req{delta-int}.  The stars are
obtained from measurements of the proton structure function and
total $\gamma^* $ p cross section \cite{Adloff:2001rw,Zeus(11)}.
References for the data on vector meson production are given in
detail in Fig. \ref{deltas}. }
\end{figure}

In Fig. \ref{all}   experimentally determined values of the power
$\delta=4\, \big(\al_{\rm P}(0)-1\big)$  for  different reactions
are displayed against the  scale $\bar b$. The values for virtual
photon scattering are deduced from measurements of the proton
structure function and the total $\gamma^* $ p cross section
\cite{Adloff:2001rw,Zeus(11)}. The experimental $\delta$ values
for vector
 meson production are taken from: a) $\rho$-production
\cite{Zeus(98),Zeus(99),Zeus(07),H1(00),H1(10)};
 b) $\phi$-production \cite{H1(10),Zeus(05)};
c) $J/\psi$-production \cite{H1(06),Zeus(02),Zeus(04),H1(13)}; d)
$\Upsilon$-production \cite{Zeus(09),LHCb(15)}. They are given  for fixed
$Q^2$, and the corresponding scale $\bar b$ has been determined by
Eq.\req{fit} with the constants collected in Table \ref{res}. The dashed
line corresponds to the fit with Eq.\req{laga} to the photon data with
\begin{equation}
  Q^2=2.354^2/{\bar b}^2+0.005~.  
\end{equation}
The solid line includes the shrinkage
correction, Eqs.\req{shrink1a},\req{shrink3},\req{delta-int}, 
 to be applied for the integrated cross sections of diffractive vector
meson production.

The errors in vector meson production and corresponding  
fluctuations are generally quite large, but the figure shows that
the data  are well compatible with  a common power behaviour, only
dependent on the $\bar b $ scale, but not on the process.
 Future data in the TeV region with reduced errors may  provide
decisive tests for the conjecture of a single Pomeron with a scale-
dependent intercept governing the energy behaviour universally for
all diffractive processes. A detailed comparison with experimental 
data is presented in Sect.~\ref{data1}.

\section{Energy dependent dipole cross section \label{edip}}
\begin{figure}
\begin{center}
\includegraphics*[width=7cm]{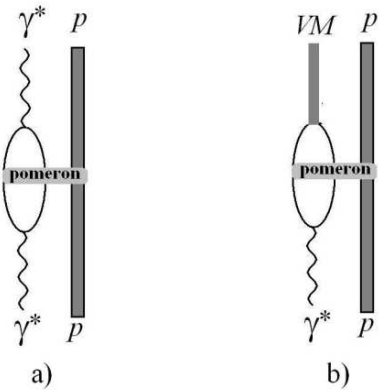}
\end{center}
\caption{ \label{dipole} Electromagnetic diffractive processes in
the dipole model with pomeron exchange. During the interaction
time the photon polarizes into a quark-antiquak pair. \quad
a: ~$\ga^*$ scattering; b: ~electroproduction of vector mesons}
\end{figure}

In this section we  explore another way to accommodate the
observed energy behaviour of electromagnetically induced
diffractive processes. As mentioned in Sect. \ref{intro}, two
Regge poles can describe  the energy dependence in the HERA range
of energies (up to 300 GeV) very well.  
It is evident that with a larger number of poles or a by
introducing a Regge cut one can extend the range of agreement to a
larger energy interval. This situation is simulated in the framework of
the dipole model  ~\cite{Nikolaev:1991et,Ewerz:2006vd,Ewerz:2007md},
which relates electromagnetic  processes  to purely hadronic
interactions through the assumption that the photon-hadron
interaction occurs via the interaction of the target hadron with a
quark-antiquark pair, as illustrated in  Fig. \ref{dipole}. This
approach has been tested in many analyses and applications.
Although there are certain limitations, 
\cite{Ewerz:2006vd,Ewerz:2007md} it is intuitive and
phenomenologically very successful.

In   the dipole model the  scattering amplitude is determined  by
a dipole cross section $\si_{\rm dip}(b_\perp, u)$, which
describes the interaction of a proton with a quark-antiquark pair
with geometrical separation $b_\perp$ and longitudinal momentum
fractions $u$ and $(1-u)$ respectively,  and the overlap functions
$\rho$ are given by the expressions \req{rhotr},\req{rhol},\req{vmtr},
and \req{vml}. 

 The forward scattering amplitude is generically given by the expression
  \beqa
\label{amp}
&& \cT_{0,\rm{pol}}= i W^2 \,  \int_0^\infty d b_\perp  \,  \\
&& \int_0^1 du\, {b_\perp }\, \sigma_{\rm{pol}}(b_\perp,u,W) ~ \rho_{\rm{pol}}(Q^2,u,b_\perp) ~ . \nonumber
\enqa 
For the energy dependence of the cross section we make the ansatz 
\beq \label{dipw}
 \sigma_{\rm{pol}}(b_\perp,u,W)  =  \sigma_{\rm dip}(b_\perp,u) (W/W_0)^{2 \, \be_{\rm{pol}}(b_\perp,u)} ~.
\enq 
The energy-dependent scattering amplitude is, thus, given by
  \beqa
\label{ampw} \cT_{0,\rm{pol}}&=& i W^2 \,  \int_0^\infty d b_\perp  \,
\int_0^1  du\,
{b_\perp }\, \sigma_{\rm dip} (b_\perp,u) \times\\
&& \qquad (W/W_0)^{2 \, \be_{\rm{pol}}(b_\perp,u)}\,  \rho_{\rm{pol}}(Q^2,u,b_\perp) ~ . \nn 
\enqa
Here again  the index ``$\rm{pol}$''  refers to longitudinal and transverse polarizations. The total 
cross section is the sum of the longitudinally and transversely polarized forms. 

Before we determine the power $ \be_{\rm{pol}}$ from experiment, we shortly
discuss  some general features of its energy behaviour. Since the
effective power of the energy   increases with increasing
photon virtuality, the function $\be_{\rm pol}(b_\perp,u)$ in Eq. (\ref{dipw})  must 
monotonously decrease with increasing $b_\perp$ and approach a value near
0.09 if the separation $b_\perp$  approaches the confinement scale (hadronic values). It is also
very plausible that there exists a value   $\be_{\rm
max}= \max [ \be_{\rm pol}(b_\perp,u)]$. In the 
limit  $W \to \infty$ the
main contribution to the amplitude \req{ampw} comes from
a region where  $\be(b_\perp,u) $ reaches its maximum and
therefore is driven to the behaviour $\cT_0 \sim (W/W_0)^{\be_{\rm
max}}$. In this respect the model seems to correspond to a Regge
cut with branch point at $\ell =\be_{\rm max}$. However, the details of the
high energy behaviour,  depend crucially on the special
form of the function $\be_{\rm pol}(b_\perp)$. In Appendix 2 we discuss the behaviour analytically.

From perturbation theory it is known that for small distances the dipole cross 
section behaves like $\si_{\rm dip}\sim b_\perp^2$, while for large distances 
it is model-dependent. We  have studied 
two different cases, namely
 \beq \label{casea}
\sigma_{\rm dip}(b_\perp,u)=\left\{ \begin{array}{ccc} C \, b_\perp^2 &\mbox{ for }& b_\perp \leq b_c\\
 C \,b_c^2& \mbox{ for }& b_\perp > b_c
\end{array}\right.
\enq and
 \beq \label{caseb}
\sigma_{\rm dip}(b_\perp,u)=\left\{ \begin{array}{ccc} C \, b_\perp^2&\mbox{ for }& b_\perp \leq b_c\\
2 C \,b_c(b-b_c) & \mbox{ for }& b_\perp > b_c
\end{array}\right.
\enq 
with $b_c = 5 ~ \rm{GeV}^{-1}$, and found that the difference on
the energy behaviour is  very small. We have used in the following  the cross section, Eq. \req{caseb}.  
Comparison 
of the amplitude $\cT_0$, Eq. \req{amp},  with the function  $ Y(Q^2,b_\perp)$, Eq. \req{Y},
shows that for $b< 5 $ GeV$^{-1}$ the amplitude $\cT_0$ receives
its main contribution from the region of $b_\perp$ where the
function $ Y(Q^2,b_\perp)$ is maximal. It  is therefore suggestive  to choose as
function $\be_{\rm po}(b,u)$   the right hand side of Eq. (\ref{laga3-b}).  
But it turns out that in   deep inelastic scattering the increase with
energy of the the total $\ga^*\,p$ hadronic cross section is much
too slow for all values of $Q^2 >1.5 \mbox{ GeV}^2$. The reason for
this behaviour is the slow decrease of  the function   $
Y(Q^2,b_\perp)$, Eq. \req{Y}, with increasing $ b_\perp$.
If one  chooses, however, the boost-invariant light-front separation $\zeta= \sqrt{u(1-u)}\;
b_\perp$ as a scale, i.e. if one postulates
\beq \label{dipwzeta}
 \sigma_{\rm pol}(b_\perp,u,W)  =  \sigma_{\rm dip}(b_\perp,u) (W/W_0)^{2 \, \be_{\rm pol}(\zeta)},
\enq 
one obtains with the functions
 \beqa\label{laga3-z}
\tilde \be_T(\ze) &=& 0.0481 \log\left[\frac{10.47}{\ze^2} +6.541\right] \\
\tilde \be_L(\ze) &=& 0.0481 \log\left[\frac{17.68}{\ze^2}
+6.530\right] \nn \enqa
 for the total $\ga^*\,p$ cross section (structure functions) very satisfactory results.  
 These expressions are derived from the maximum  $\zeta(Q^2)$  of the
function 
\beqa\label{Z}
 Z(Q^2,\zeta)&=& \int_0^1 du \,\si_{\rm dip}\left(\frac{\ze}{u(1-u)},u\right) \times\\
&&   \rho_{\ga^*\ga^*, \rm
pol}\left(Q^2,u,\frac{\zeta}{u(1-u)}\right) ~,  \nn \enqa 
and the relation between $Q^2$ and the power behaviour given by Eq. (\ref{laga}).
Here the energy behaviour cannot be described by a single power and therefore one has to fit 
 effective powers for a certain energy range. From the results obtained with 
Eqs. (\ref{ampw},\ref{dipwzeta},\ref{laga3-z})  the theoretically
obtained curves in the range of energies 20 -- 200 GeV
 (HERA range) we obtain the results displayed in Table \ref{gam-z}, which 
compare favorably with the data \cite{Adloff:2001rw,Zeus(11)}.
 We give also the theoretical effective power fitted  in the energy range  200 GeV - 2 TeV 
(accessible at LHC). As can be seen, the difference  of values is  
only  of about 10 \%.
\begin{table}
\begin{center}
\begin{tabular}{|c|cc|c|}
\hline
$Q^2$& \multicolumn{2}{|c|}{$\lambda$(theory)}&$\lambda$(experiment) \\
GeV$^2$&HERA&LHC& HERA\\
\hline
2&0.183&0.205&0.159$\pm$0.016\\
5&0.213&0.237&0.196 $\pm$ 0.01\\
15&0.258&0.288&0.250 $\pm$0.01\\
25&0.280&0.310&0.274 $\pm$ 0.015\\
45&0.306&0.335&0.302 $\pm$ 0.02\\
60&0.319&0.348&0.332 $\pm$ 0.026\\
90&0.337&0.365&0.304 $\pm$ 0.05\\
\hline
\end{tabular}
\end{center}
\caption{\label{gam-z} Values of the effective power $\la=
\alp-1$ for the total $\ga^*\,p$ cross section (structure
function). Experimental values are from
\cite{Adloff:2001rw,Zeus(11)}, the theoretical values are obtained
with the energy-dependent dipole cross section \req{ampw} and the
expressions \req{laga3-z}. The HERA column indicates that  the effective power
is fitted in the HERA range of energies (ca 40 -200 GeV), whereas
LHC is the theoretical value in a range accessible for LHC (200
GeV to  2 TeV).}
\end{table}

The cross section obtained with Eqs. \req{ampw} and \req{laga3-z} can be well parameterized 
by an energy-dependent power,
\beq
\si_{\rm int} = C (W/W_0)^{\delta(W)}
\enq
with 
\beq
\delta(W) = E + \frac{F}{1000} \log[W/W_0]\enq
The constants $E$ and $F$ are given in Table \ref{tab-par}.

%%%%%%%%%%%%%%%%%%%%%%%%%%%%%%%%%%%%%%%
\section{Description and prediction of diffractive vector meson production  \label{data1} }

In this section we compare   data and predictions of the models discussed in Sects. \ref{scale1}
and \ref{edip}.

\begin{figure*}
\begin{center}
\includegraphics*[width=8cm]{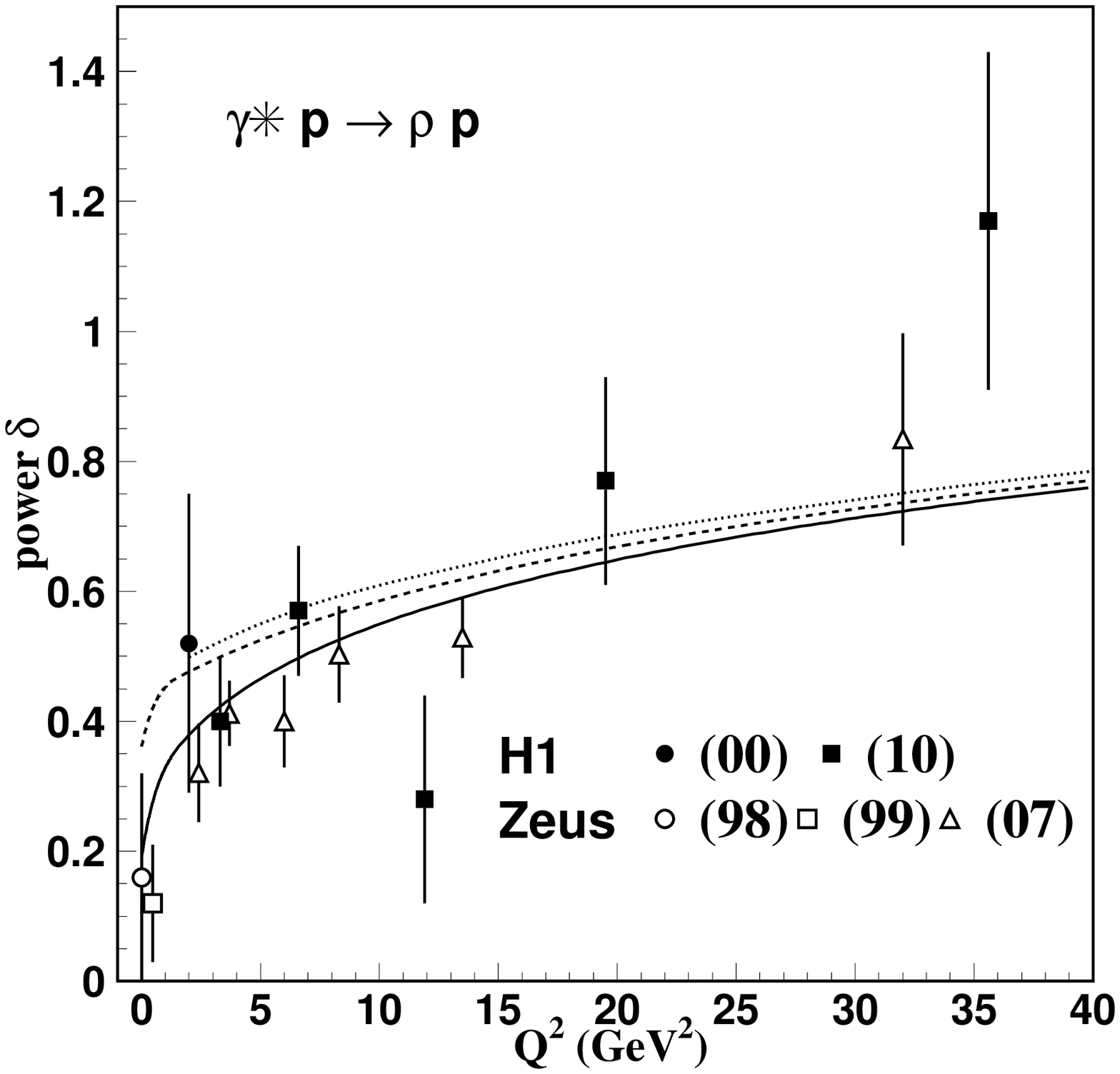}
  \includegraphics*[width=8cm]{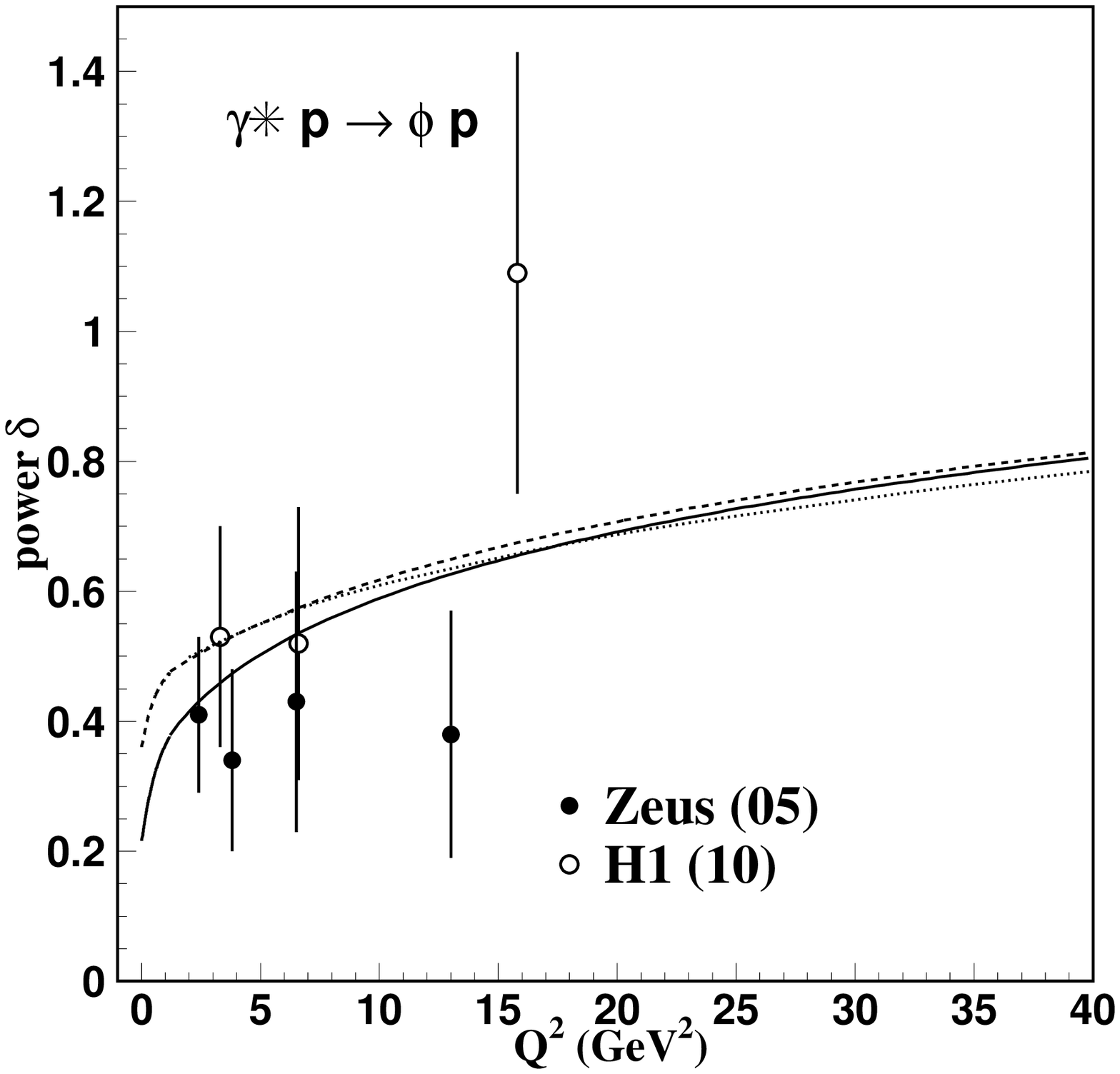}
  \includegraphics*[width=8cm]{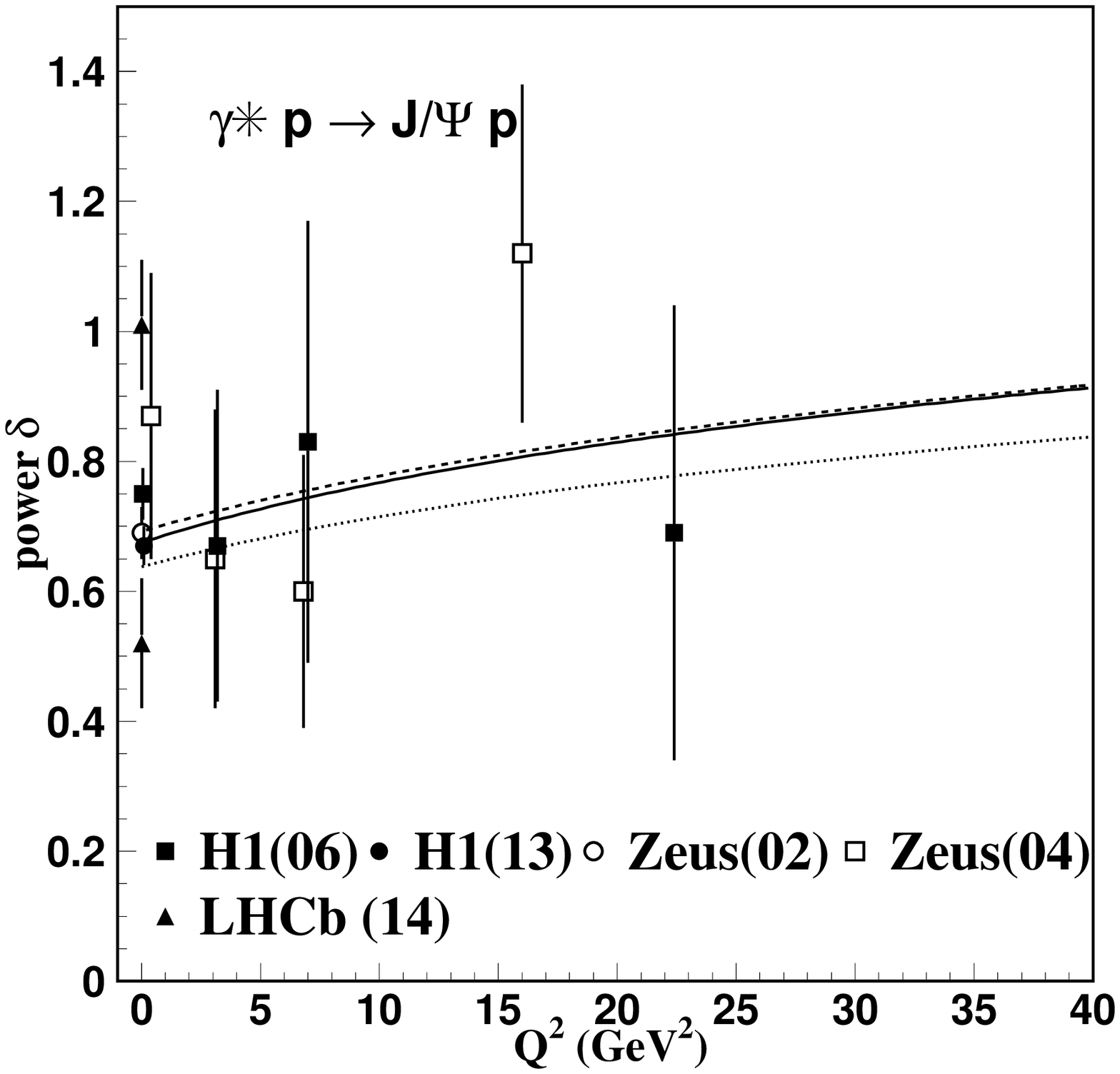}
\includegraphics*[width=8cm]{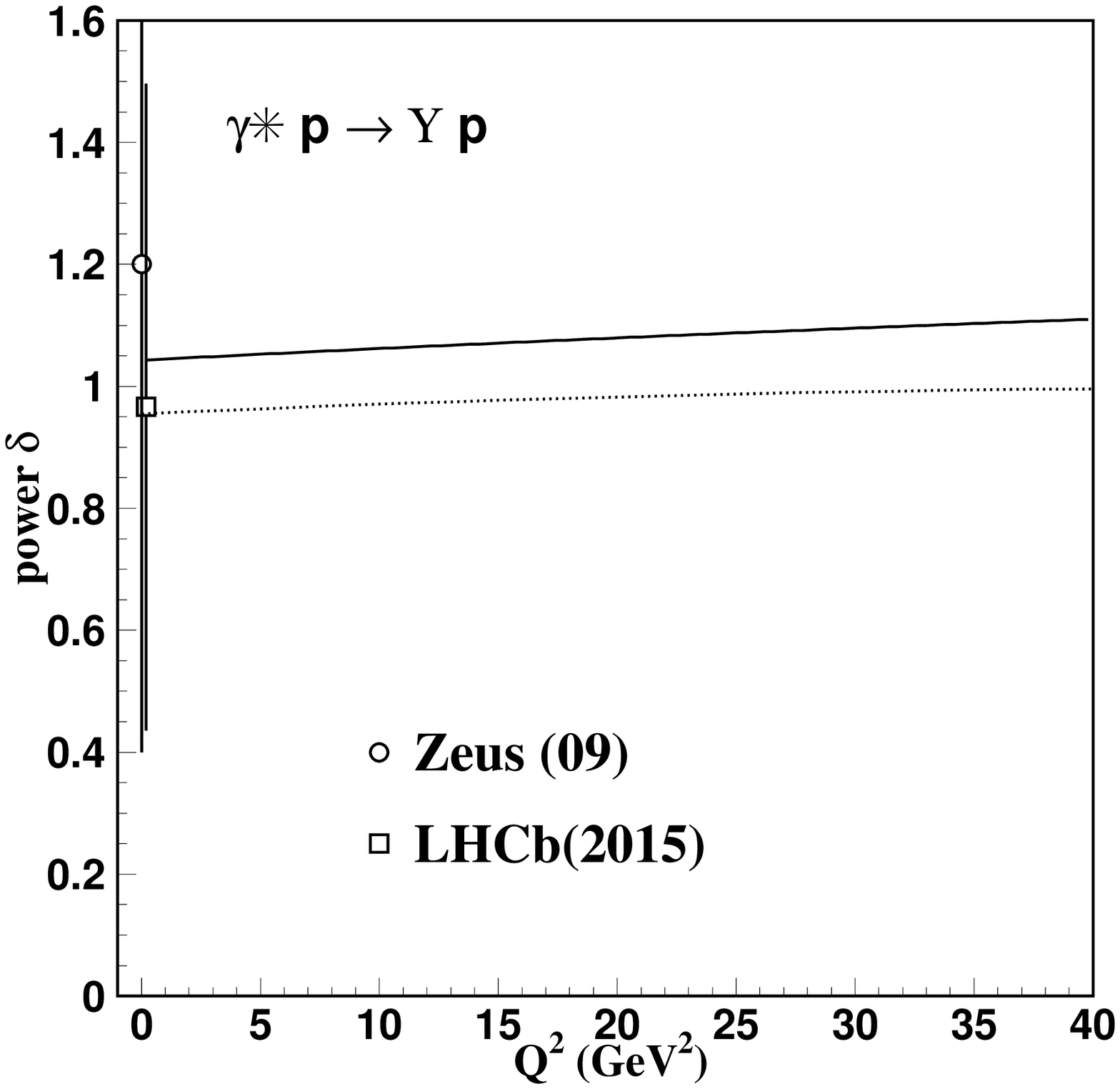}
\end{center}
\caption{ \label{deltas}Predictions for the power $\delta$ and
$\delta_{\rm int} $   as function of $Q^2$ for vector meson
electroproduction  and experimental values for this quantity:  a)
$\rho$-production \cite{Zeus(98),Zeus(99),Zeus(07),H1(00),H1(10)};
 b) $\phi$-production \cite{Zeus(05),H1(10)} ;
c) $J/\psi$-production \cite{H1(06),H1(13),Zeus(02),Zeus(04)}; d)
$\Upsilon$-production \cite{Zeus(09),LHCb(15)}.
 The solid  line and dashed lines represent respectively
$\delta_{\rm int}$, Eq.\req{delta-int}, including  the shrinkage
correction, and $\delta$, Eq. \req{meq}, without shrinkage
correction. The dotted lines are  the results of the effective power 
 as obtained from the model with an energy-dependent dipole cross section discussed 
in Sect. \ref{edip},    Eq. (\ref{ampw})  with  
the energy dependence given by Eq. (\ref{laga3-z})  without shrinkage 
correction.  }
\end{figure*}

In Fig. \ref{deltas} and Table \ref{del}  we display  theoretical predictions  
 for the powers $\delta$ and $\delta_{\rm
int}$, that is,  without and with shrinkage correction,  as 
functions of the photon virtuality $Q^2$  for unpolarized elastic
production of  vector mesons in the ground state, together with the experimental results. The
theoretical results for the  $\omega$ meson production are not
distinguishable from those of $\rho$ production. The long-dashed and the solid lines are obtained 
with the model discussed in Sect.\ref{scale1};  the long-dashed
curves represent the uncorrected power $\delta(Q^2)$, obtained
from Eq. (\ref{meq}), and the solid line represents $\delta_{\rm int}$,
Eq. (\ref{delta-int}), that includes  shrinkage corrections.  The dotted line is the result of the model 
discussed in Sect. \ref{edip}, where an effective power $\delta$ has been extracted from the 
energy range  $20 \leq W \leq 200$ GeV. The shrinkage corrections to these results are
the same as those for the results of  Sect. \ref{scale1}.
The theoretical values of $\delta(Q^2)$  calculated for $Q^2 \geq 1$ GeV$^2$ and extrapolated to the 
value 0.36 at $Q^2 =0$.  The
theoretical predictions of both models are well  compatible with the data. The
observed  sharp increase of the power delta with $Q^2$ near
$Q^2=0$ for the light vector mesons indicates that the rapidly
varying shrinkage correction given by Eq. (\ref{shrink3}) is quite
realistic. As can be seen, the shrinkage corrections are only important for 
$\rho$ production  at $Q^2 \leq 10 \; \rm{GeV}^2$. 

\begin{figure*}
\begin{center}
\includegraphics*[width=7.2cm]{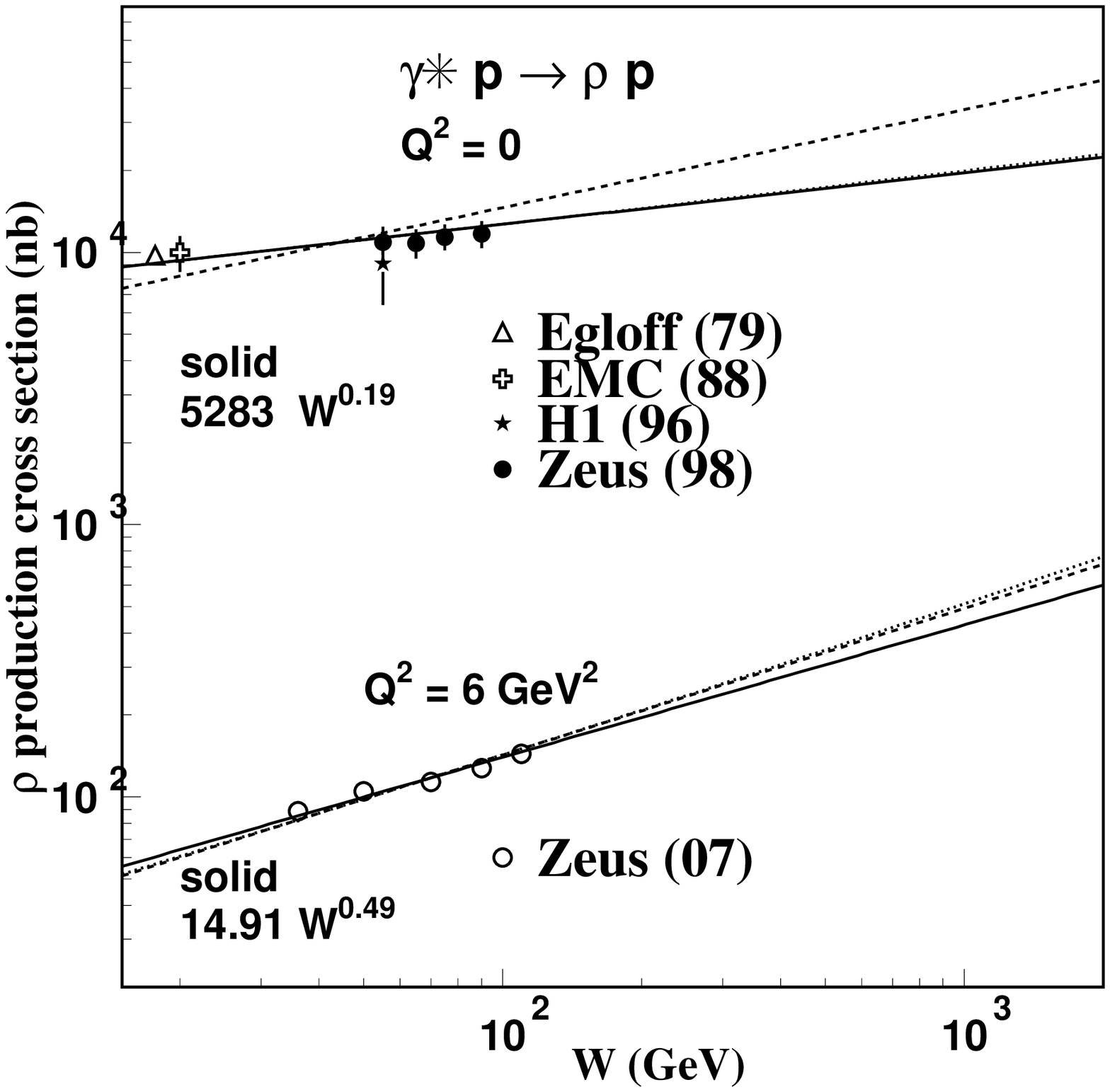}
\includegraphics*[width=7.2cm]{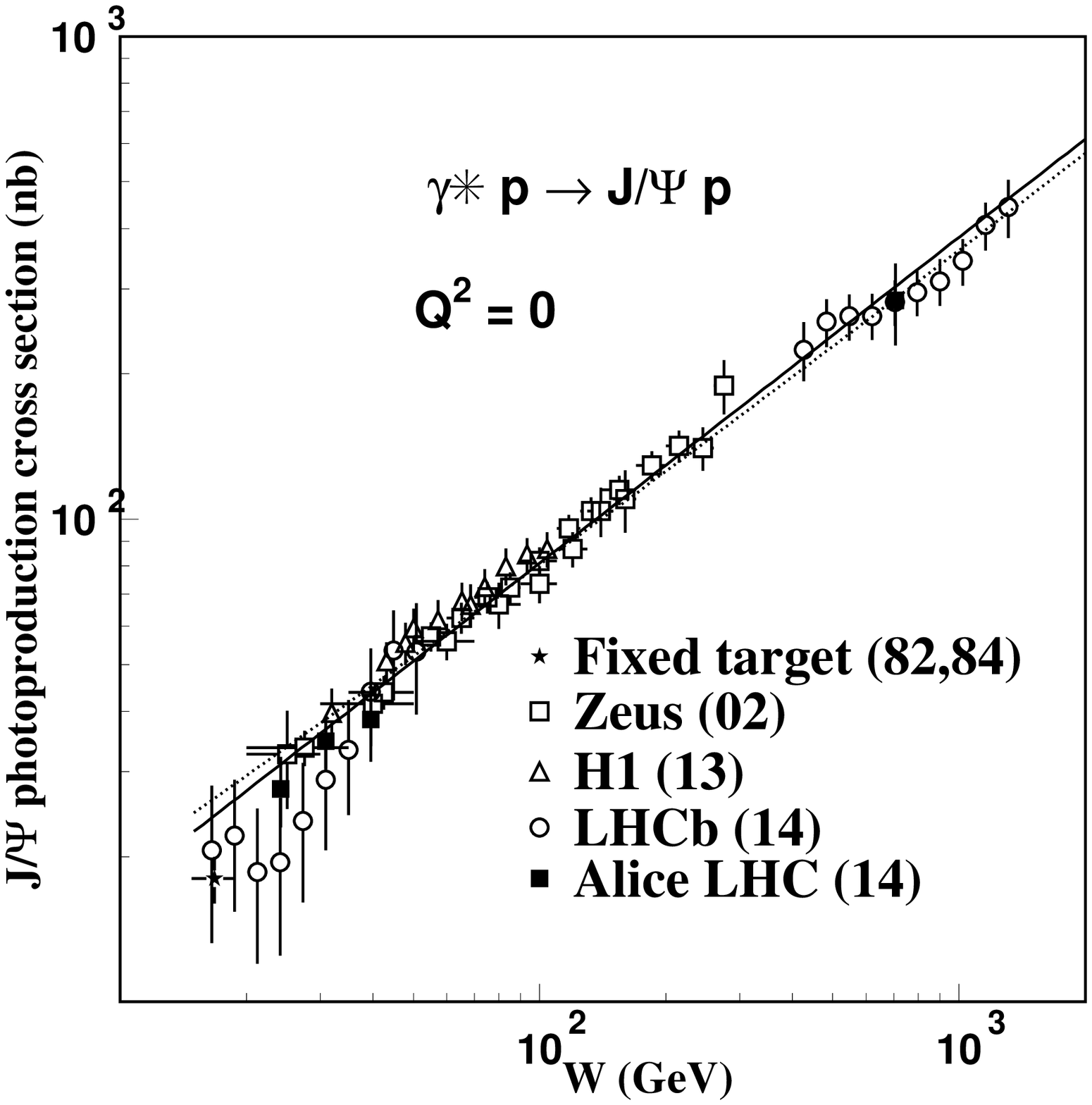}
\includegraphics*[width=7.2cm]{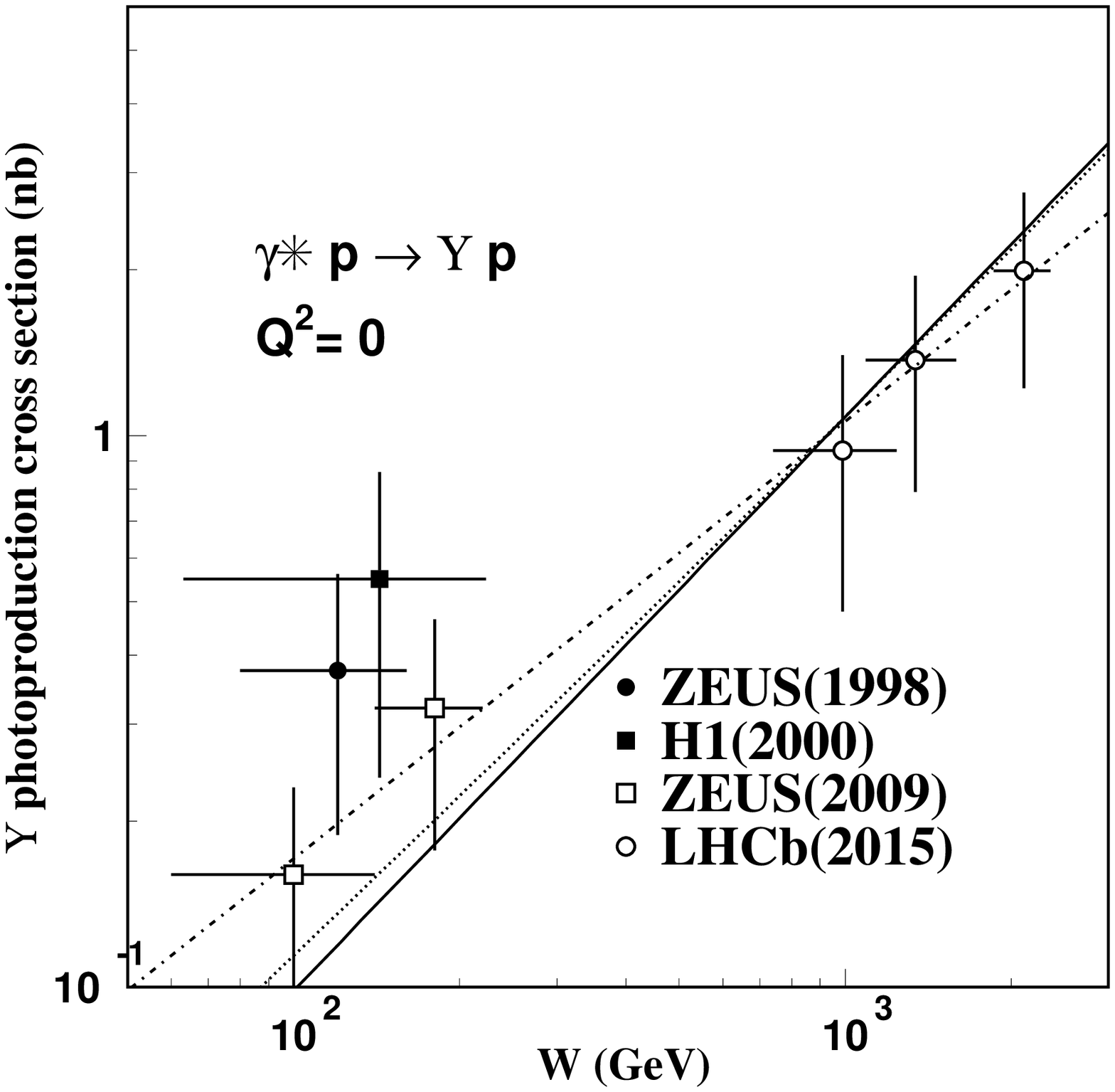}
\end{center}
\caption{\label{cs} Energy dependence of photoproduction cross sections. 
   In $\rho$ production \cite{Egloff,EMC,H1(96),Zeus-PL98,Zeus(07)},   
the results of Sect. II,  with a scale-dependent 
   Regge singularity, and those of Sect. III, with energy-dependent 
   dipole cross section, are practically indistinguishable; 
   the dashed line is the result without shrinkage correction, Eq.(19), and the 
   solid includes the shrinkage correction, Eq.(25). In $J/\psi$  
   \cite{Fixed_Target,Zeus(02),H1(13),LHCb(14),Alice(14)}  and $\Upsilon $
   \cite{Zeus(98),H1_PLB2000,Zeus(09),LHCb(15)}
     production, the shrinkage corrections are negligible. The solid lines 
   are the result of Sect. II and the dotted line is the result of section 
   III. For $ J/\psi$  production a free fit coincides with the result of 
   Sect. II. For $\Upsilon$ production it is shown with a dot-dashed 
   line (the two old points in full circle and full square are excluded 
   in this fitting). The precise parameters for the curves are given 
   in Table \ref{tab-par}.}
\end{figure*}

In    Fig.  \ref{cs}  data and  the theoretically predicted energy
dependence of $\rho$, $J/\psi$  and $\Upsilon $  production cross sections are 
displayed.
In  the model of Sect. \ref{scale1} the energy dependence is represented by a
single power  $C\,W^\delta$ in the full energy range. Since emphasis in this 
paper is on energy dependence and the absolute values of the cross sections 
depend  on details of the models, the constant
$C$ is fitted to the data, and the values for the power $\delta$ are
given by the model, both without  shrinkage corrections, i.e. $\delta$ 
from Eq. (\ref{meq}) (dashed lines) and with shrinkage corrections, 
Eq. (\ref{delta-int}), solid line. The dotted lines are results of the dipole model of 
Sect. \ref{edip}, including shrinkage corrections according to Eq. (\ref{shrink3}).

The plot of $J/\psi$ photoproduction  
includes the most recent LHC data \cite{LHCb(14),Alice(14)}.  Here the influence of the
shrinking correction is very small:  $\delta(0) = 0.69$ and
$\delta_{\rm int}(0) = 0.67$. The
fit of the 58 points with  free power gives the same value 0.67.
Very recent values for $\Upsilon$ production together with theoretical predictions 
are also shown. Within errors they are compatible with both models. 

In Table \ref{tab-par} we have collected the parameters of the theoretical curves 
displayed in Fig. \ref{cs}, together with values of unconstrained fits to the data. 
 %   [GeV$^2$]
%%%%%%%%%%%%%%%%%%%%

The transverse and longitudinal wave functions are different and
therefore we obtain  different scales for the respective cross
sections. This leads to a different energy behaviour for the two
polarizations. According to the scale-dependent Regge pole, as discussed 
in Sect. \ref{scale1}  the ratio $R=\si_L/\si_T$ has the power
behaviour 
\beq \label{longtrans} R= \frac{\si_L}{\si_T}= A\,
W^{\delta_{\rm R}}~, \enq with \beq \label{lt2} \delta_{ R} =
\delta_L- \delta_T ~ . \enq 
The values of $\delta_L$ and
$\delta_T$     are determined by Eq.\req{meq} with the constants
$A_{\rm VM,long},\, A_{\rm VM,trans},\,B_{\rm VM,long},\,B_{\rm
VM,trans}$ of Table \ref{res}. In the model with an energy-dependent dipole cross section, see Sect. \ref{edip},
 the corresponding expressions are obtained by calculating separately the longitudinal and transverse cross sections with the power functions  (\ref{laga3-z}).

\begin{figure*}[h]
\begin{center}
\includegraphics*[width=8cm]{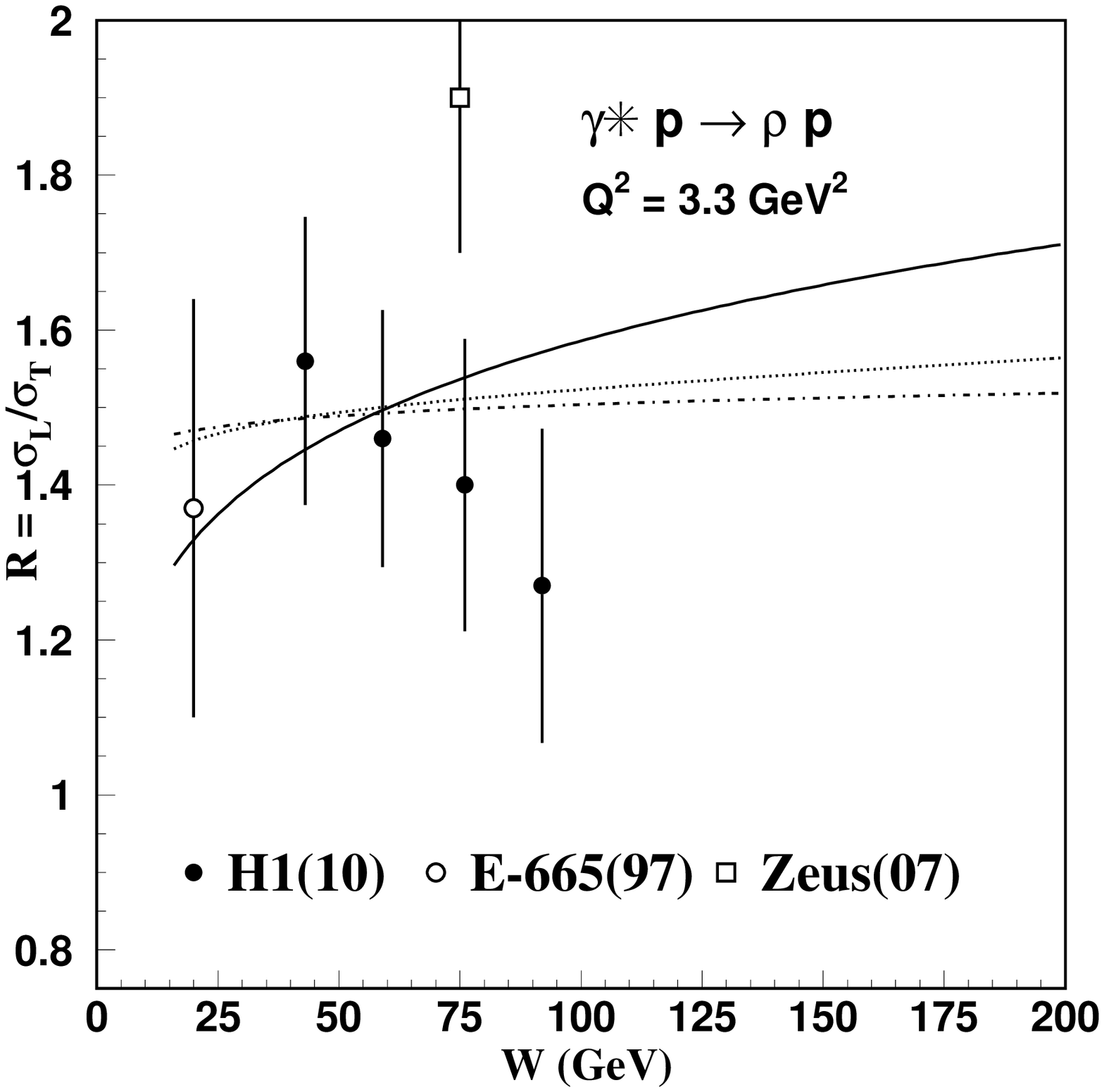}
\includegraphics*[width=8cm]{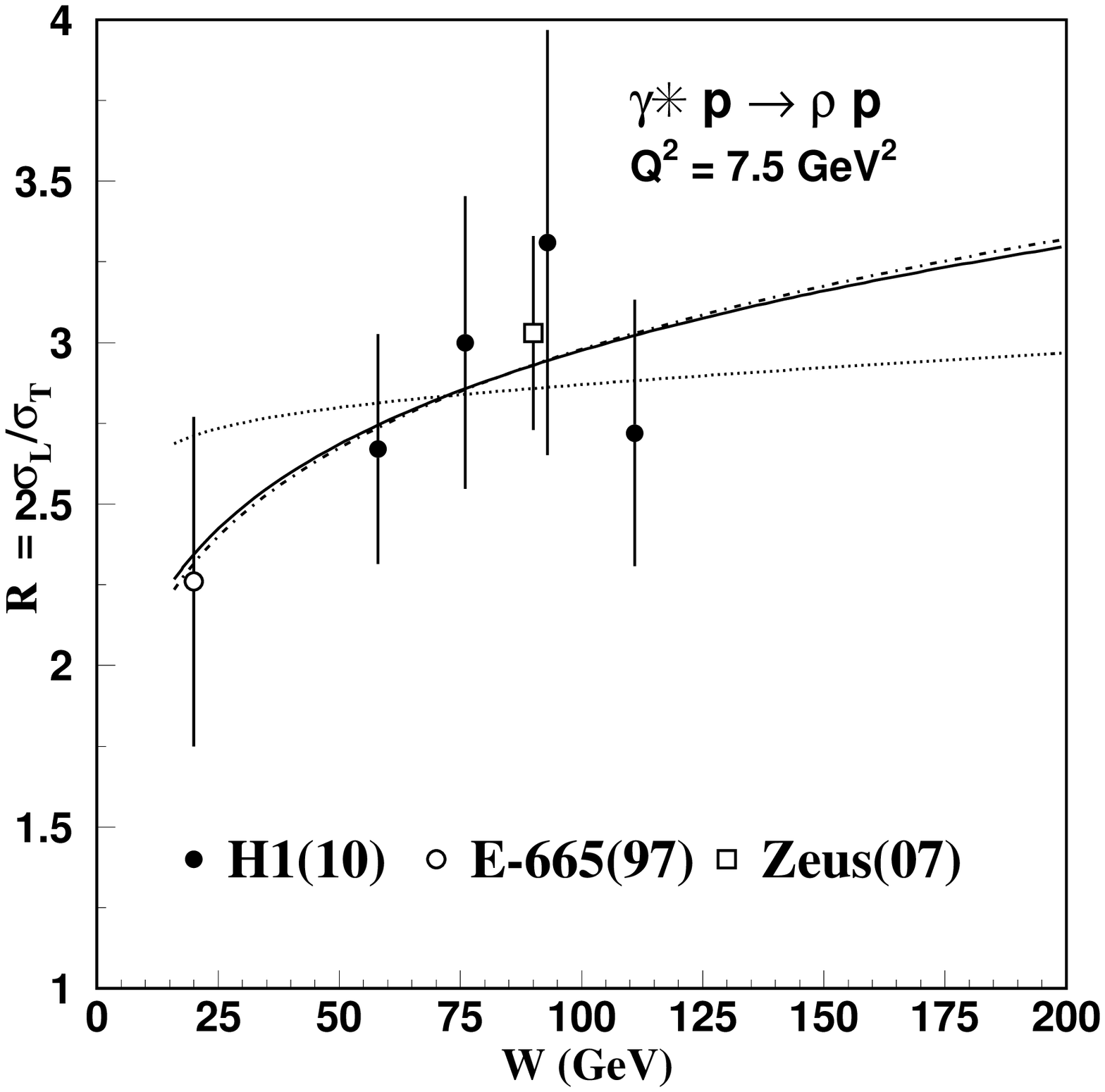}
\includegraphics*[width=8cm]{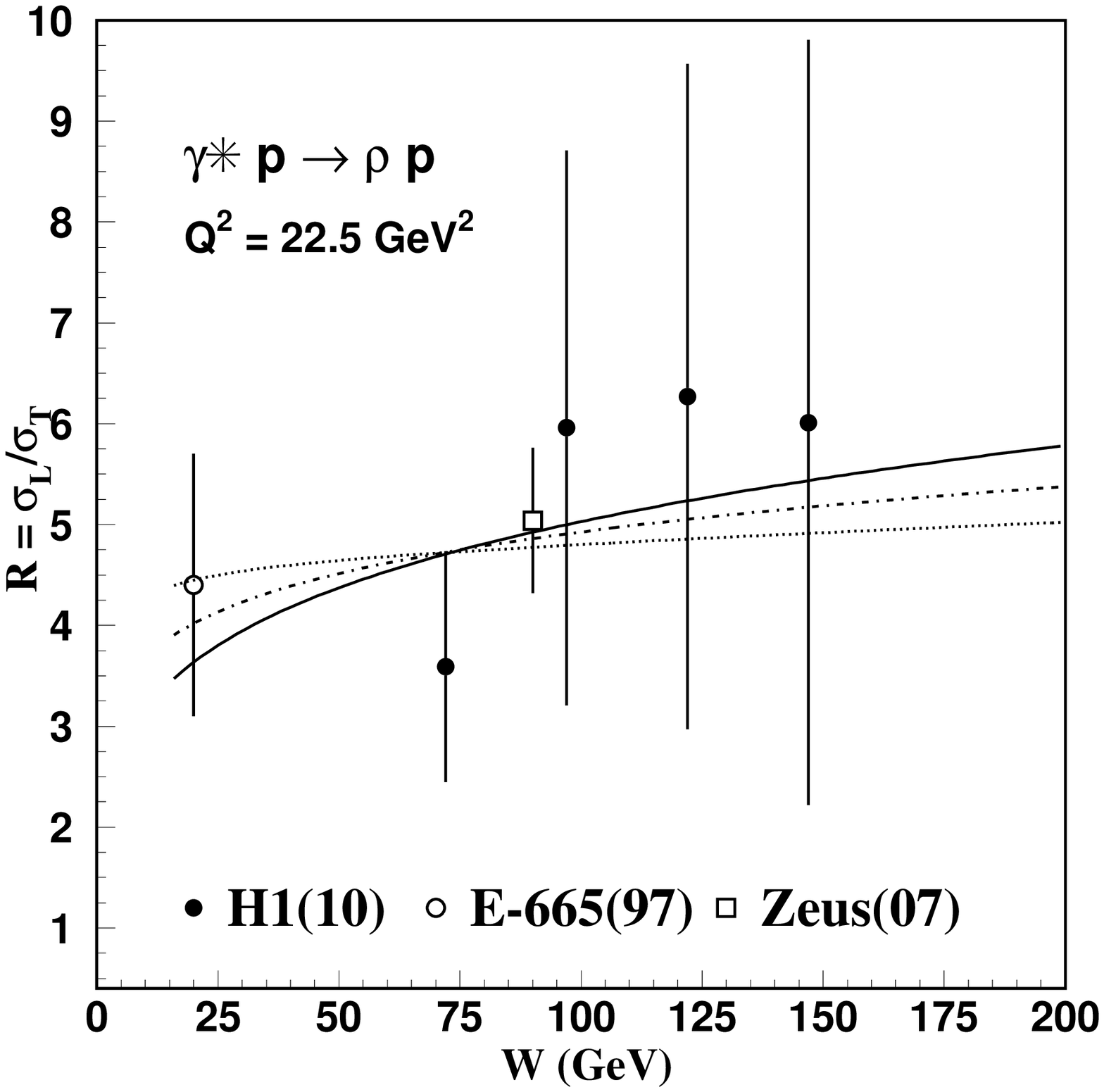}
\includegraphics*[width=8cm]{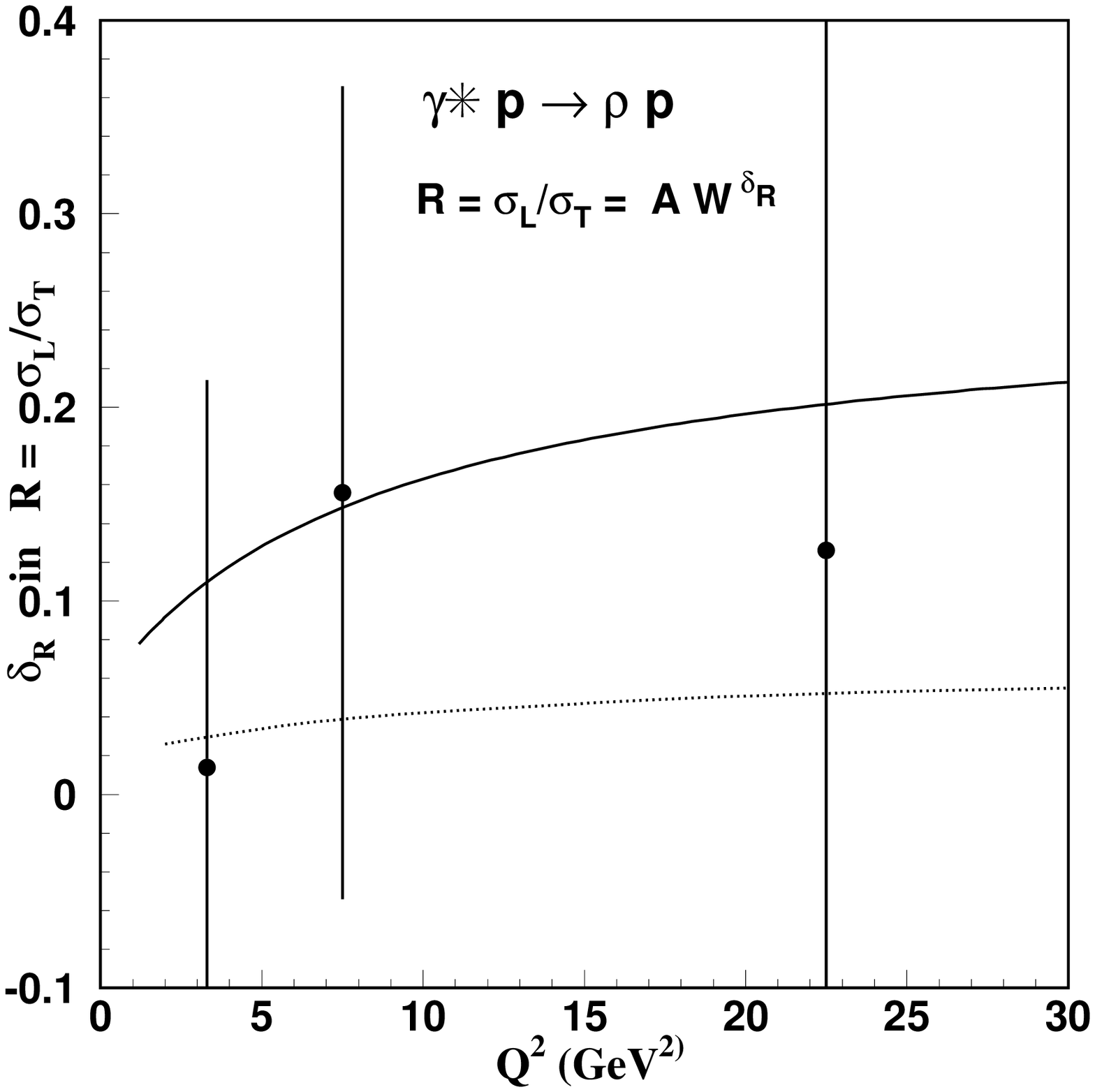}
\end{center}
\caption{ \label{ratios-fig} a)-c): Energy dependence of the
ratio of longitudinal to transverse cross section of $\rho$
electroproduction for $Q^2=3.3, 7.5$ and 22.5 GeV$^2$. The solid
lines have the energy dependence predicted according to
 Eq.\req{longtrans} with   Eq.\req{meq}, 
  the dotted lines the results of Sect. \ref{edip}, Eqs.\req{ampw},\req{laga3-b}, 
and  the dot-dashed   lines are free best fits to the data.  
The data are from
\cite{Zeus(07),H1(10),E665(97)} ; d) Power $\delta_R$ for the energy
dependence of the ratio as function of $Q^2$. The solid line is
$\delta_L-\delta_T$ from Eq.\req{meq}, and the points with error bars
are from the best fits to the data presented in the plots a)-c).}
 \end{figure*}

In Fig. \ref{ratios-fig} a) - c) we show  data
\cite{Zeus(07),H1(10),E665(97)}
 for the energy dependence of the polarization ratios $R=\si_L/\si_T$
for $\rho$ production at three values of $Q^2$. The solid lines are the theoretical
predictions according to Sect.\ref{scale1}, Eqs. (\ref{meq},\ref{longtrans},\ref{lt2}).
The multiplicative constant $A$ in Eq. \req{longtrans} is fitted
freely. The dotted lines are the results of the energy-dependent dipole model, Sect. \ref{edip}. 
 We also show in dot-dashed lines the results of  free fits to
the data with  unconstrained $A$ and $\delta_R$. At $Q^2=$  7.5
and 22.5 GeV$^2$ the  model gives good agreement for the energy
dependence of $R$. In the last plot of the set, the data
and the theoretical predictions   for the power coefficients as
functions of $Q^2$ are compared directly.
%%%%%%%%%%%%%%%%%
%%%%%%%%%%%%%%%%%%%%%%%%%

The experimental errors for the
ratio $R$ are quite large and also the theoretical uncertainties
in the small differences between $\delta_L$ and $\de_T$ are
large and both models are compatible with the data. The numerical 
values are given in Table \ref{b-z}.

\begin{table}
\begin{center}
\begin{tabular}{|c|ccc|cc|}
\hline
$Q^2$& \multicolumn{3}{|c|}{$\de_{\rm tot}$}&\multicolumn{2}{|c|}{$\de_L-\de_T$} \\
GeV$^2$&Regge&\multicolumn{2}{c|}{Dipole}&Regge&Dipole\\
&&HERA&LHC&&HERA\\
\hline
\multicolumn{6}{|c|}{$\rho$}\\
2&0.476&0.499&0.512&0.091&0.026\\
5&0.525&0.550&0.569&0.128&0.034\\
15&0.632&0.651&0.682&0.183&0.047\\
30&0.727&0.741&0.779&0.213&0.055\\
45&0.790&0.802&0.843&0.226&0.058\\
\hline
\multicolumn{6}{|c|}{$J/\psi$}\\
0&0.692&0.638&0.663&0.067&0.024\\
5&0.740&0.681&0.711&0.078&0.026\\
15&0.809&0.743&0.778&0.090&0.027\\
30&0.881&0.806&0.844&0.100&0.028\\
45&0.934&0.851&0.894&0.105&0.030\\
\hline
\multicolumn{6}{|c|}{$\Upsilon$}\\
0&1.046&0.955&0.995&0.060&0.013\\
15&1.073&0.977&1.020&0.062&0.013\\
\hline
\end{tabular}
\end{center}
\caption{\label{b-z} Theoretical values for the effective power
$\de$ of the energy dependence $W^\de$ for integrated vector meson
production  cross section ($\de_{\rm tot}$) and   for  the
ratio of  of longitudinal and transverse polarized cross section
($\de_L-\de_T$). The columns marked "Regge" show the results of a
scale dependent Regge pole as discussed in Sect. 
\ref{scale1}, those under "Dipole" are the results of the
energy-dependent dipole cross section treated in Sect. \ref{edip}; 
the columns HERA indicate  the effective power fitted in the HERA range of
energies (ca 40 -200 GeV), while  LHC gives  the  value in a range
accessible to LHC (ca 200 GeV - 2 TeV).}
\end{table}

\section{Summary and Conclusions  \label{final}}

We  have presented two simple   phenomenological models which
account for two striking features of electromagnetic diffractive
processes, namely that the energy behaviour can be well described
by a power behaviour and that the power parameter  increases with increasing
photon virtuality $Q^2$. It is remarkable that the power behaviour
observed in elastic diffractive $J/\psi$ photoproduction in the
HERA range of energies describes the data also up to 2 TeV. Such a
behaviour is natural in Regge theory. In order to describe the
observed dependence on the photon virtuality by a single pole we
have to assume, however, that the position of this pole in the
complex angular momentum plane depends on $Q^2$ for negative squared
momentum transfer $t$ . In Sect. \ref{scale1} we have
shown that this behaviour is not in contradiction with general
principles and gave the prescription for calculating the position of 
the scale-dependent Regge pole. 

In Sect. \ref{data1}  it was shown that the model is very well compatible 
with experiment,  and very recent 
data on $\Upsilon$ production at LHC  \cite{LHCb(15)}  confirm it further.  
Also the concept 
of a scale-dependent slope of the Pomeron
trajectory ~\cite{Brower:2006ea} is well compatible with the data
\cite{H1(10)}, as shown in Fig.~\ref{cs}.

\begin{figure}[h!]
\begin{center}
\includegraphics[width=7.5cm]{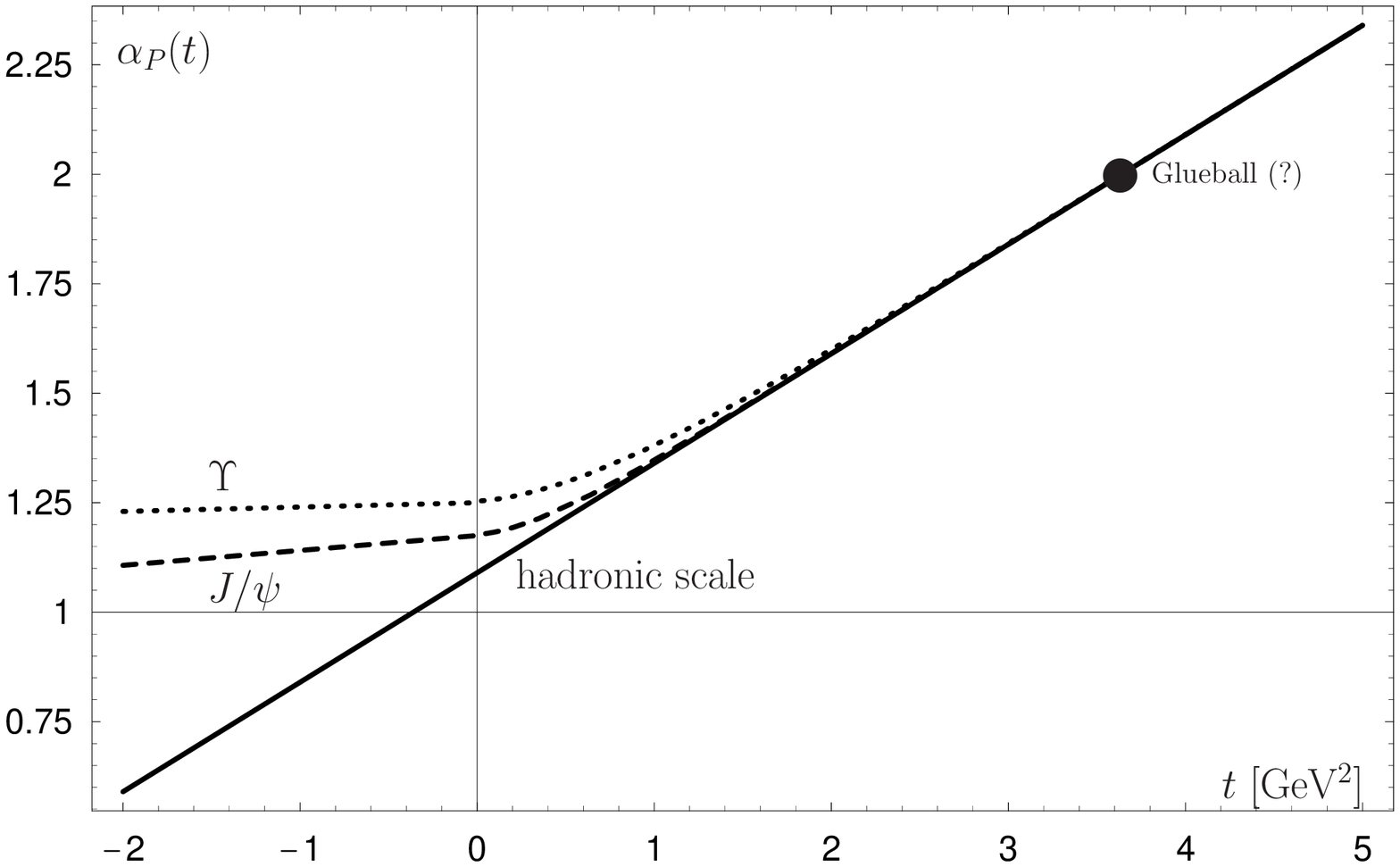}
\end{center}
 \caption{\label{traj} Scenario for scale-dependent Pomeron trajectories:
the solid line is the  trajectory relevant at hadronic scales
(soft Pomeron), and the  dashed and dotted  lines represent  the
trajectories for $J/\psi$ and $\Upsilon$ photoproduction,
respectively.}
\end{figure}

 A possible scenario for
trajectories for $J/\psi$ and $\Upsilon$  photoproduction together
with the conventional soft Pomeron trajectory  is illustrated  in
Fig.~\ref{traj}. The intercept $\al_P(0)$ and the slope for $t<0$
are fixed by the model, see Eqs.\req{meq}, \req{shrink3}. For $ t
\gg 0$,   where glueball states may be on the trajectory, the
hadronic confinement  scale becomes relevant and there it should
coincide with the soft Pomeron, that is the Pomeron trajectory
relevant for hadronic scattering.

\begin{figure}[h!]
\begin{center}
\includegraphics[width=7.5cm]{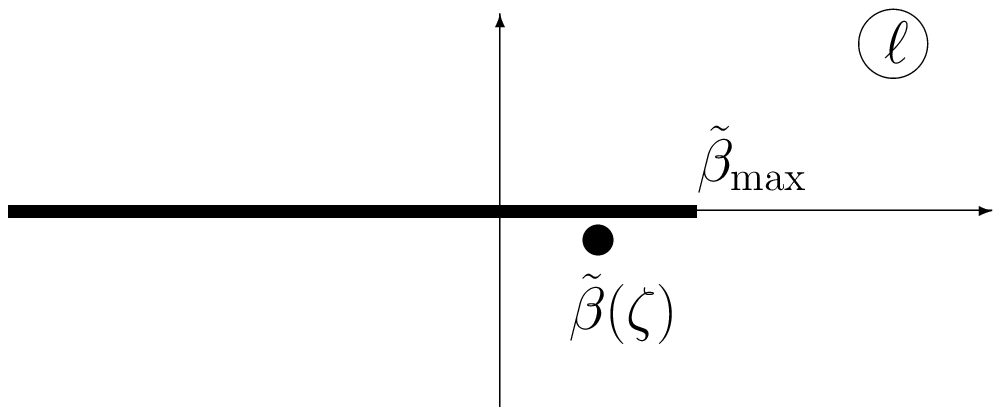}
\end{center}
 \caption{\label{cut}Scenario for a Regge cut in the angular momentum plane with a nearby pole on the second sheet. Such a situation could explain the energy-dependence obtained from the energy-dependent dipole cross section discussed in Sect. \ref{edip}. }
\end{figure}

A second model, based on a specific energy-dependence of the cross
section of a quark-antiquark dipole,
 see Eq. \req{dipw}, was  discussed in Sect. \ref{edip}. If we 
choose as relevant scale for the energy dependence of the
  dipole cross section the boost-invariant light-front separation $\zeta= \sqrt{u(1-u)}\; b_\perp$ of 
the quark-antiquark pair,
\beq \label{dipwz}
 \sigma_{\rm dip}(b_\perp,u,W)  =  \sigma_{\rm dip}(b_\perp,u) (W/W_0)^{2 \, \tilde \be(\sqrt{u(1-u)}\;  b_\perp)} ~ ,
\enq 
we also  obtain with the expressions (\ref{laga3-z}) good
agreement with the data and specifically, that the effective power
describing the energy-dependence varies only very little in the
energy range up to 2 TeV. The energy-dependence obtained in this
model  corresponds to that of a Regge cut in the complex angular
momentum plane.  The main contribution to the discontinuity across
the cut comes from the region $\ell = \tilde \be(\ze)$. This could
be realized by a pole on the second sheet near the real axis, as
indicated in Fig. \ref{cut}. For positive  values of $t$ this pole
could emerge into the physical sheet and lead to particle poles
for positive values of $t$ in the usual way.

\bigskip 
 
{\bf Note added in proof:} ~  Due to linear $t$-channel unitarity the partial wave  
amplitude for diffractive electroproduction of vector mesons, 
$\cal{T}_{\ell}$ in  Eq.(3), will in general  contain all contributions to 
the $p \bar p$ scattering amplitude, including the Pomeron at hadronic 
scales (soft Pomeron). The single power $\delta(Q^2)$ used in Sec. II is 
therefore an effective power and can deviate  from the corresponding 
value of the moving trajectory depicted in Fig. 9.  We thank Peter Landshoff 
for pointing this out.


\begin{acknowledgments}
 
 It is a pleasure to thank  Guy de T\'eramond, Carlo Ewerz, and
Otto Nachtmann for numerous constructive critical suggestions and
remarks. The author  E.F.  wishes to thank the Brazilian
agencies CNPq, PRONEX, CAPES and FAPERJ for financial support.
 
\end{acknowledgments}

\clearpage

\section{Appendix}

\subsection{Tables of Numerical Fits }
\begin{table}[h!]
\begin{center}
  \begin{tabular}{|l | c c | c c |}\hline
\multicolumn{1}{|l}{} &  \multicolumn {2}{|c|} { Transverse} &
          \multicolumn {2}{|c|} { Longitudinal}\\
&$~ \om$ & $N$ &$~ \om$ & $N$  \\
&   (GeV) & $ $    & (GeV)& $  $ \\  \hline
$\rho(770)$ &$0.2809 $  &$2.0820$&$0.3500$   &$1.8366$ \\
$\omega(782)$   &$0.2618 $  &$2.0470$&$0.3088$   &$1.8605$\\
$\phi(1020)$    &$0.3119 $  &$1.9201$&$0.3654$   &$1.9191$ \\
$J/\psi(1S)$    &$0.6452 $  &$1.4752$&$0.7140$   &$2.2769$ \\
$\Upsilon(1S)$  &$1.3333 $  &$1.1816$&$1.3851$   &$2.7694$ \\
\hline  \end{tabular}  \end{center} \vspace{-5mm}
  \caption{ \label{WFparam}  Parameters of the Brodsky-Lepage (BL) vector meson wave
functions (\ref{BL}), taken from
\cite{Dosch:2006kz,Baltar:2009vp}. }
\end{table}

\begin{table}[h!]
\begin{center}
\begin{tabular}{|c|ccc|ccc|}\hline
final state & \multicolumn{3}{|c|}{$A_{\it fs,\rm pol}$   }&\multicolumn{3}{c|}{$B_{\it fs,\rm pol}$ [GeV$^2$]}    \\
  {\it fs}  &trans&long&total&trans&long&total\\
\hline
$ \gamma^* $ &2.337&2.467&2.354&-0.003&-0.003&-0.005\\
$\rho $&10.594&5.658&7.565&3.971&2.248&2.699\\
$ \phi$&9.117&5.578&6.658&3.711&2.454&2.449\\
$J/\psi$& 6.968&5.329 &5.644&20.021&15.975&14.696\\
$\Upsilon$& 6.015&5.200&5.241&117.64&108.63&93.868\\
\hline
\end{tabular}
 \end{center}
\vspace{-5mm}
  \caption{ \label{res}
 Coefficients of the numerical fits of  the average scale
$\bar b_{\it fs} (Q^2) $ , for the processes $\ga^*
{ p} \to {\it fs} \, { p}$
 with Eq. (\req{fit}),  for use in longitudinal, transverse and total
(incoherent sum of the two cases)  cross sections. $fs=\gamma^*$
refers to $\gamma^* \, {p}$   total cross section. The accuracy of
the fit is better than 1\% in the $Q^2$  range from 1 to 60
GeV$^2$ for photon scattering and $\rho,\,\phi$  production and
from  0 to  60 GeV$^2$ for $J/\psi$ and $\Upsilon$ production.
Remark: for $Q^2$ = 0 in $\rho$ and $\phi$  production, the
relevant scale is the hadronic scale,
 chosen as $\bar b =5 \mbox{ GeV}^{-1}$, with a soft Pomeron intercept
1.09.}
\end{table}

\begin{table}[h!]
\begin{center}
\begin{tabular}{|r|c|}
\hline
Reaction  ~  ~  & $\de(Q^2)$\\
\hline
$\ga^* p \to \rho \,p, $ total&$ 0.1924\,\log (9.61827 + 1.1353\,Q^2)$\\
 trans&$0.1924\,\log (8.7997 + 0.5707\,Q^2)$\\
long&$0.1924\,\log (11.5422 + 2.22949\,Q^2)$\\
\hline
$\ga^* p \to \phi \,p, $ total&$0.1924\,\log (10.1436 + 1.46565\,Q^2)$\\
trans&$0.1924\,\log (9.3932 + 0.77065\,Q^2)$\\
long&$0.1924\,\log (12.1579 + 2.29395\,Q^2)$\\
\hline
$\ga^* p \to J/\psi \, p, $ total&$0.1924\,\log (36.5274+2.03964\,Q^2)$\\
trans&$0.1924\,\log (32.9435+1.31911\,Q^2)$\\
long&$0.1924\,\log (46.6689+2.51262\,Q^2)$\\
\hline
$\ga^* p \to \Upsilon\,  p,$ total&$0.1924\,\log (228.578 + 2.36527\,Q^2)$\\
trans&$0.1924\,\log (214.809 + 1.77039\,Q^2)$\\
long&$0.1924\,\log (293.23 + 2.63924\,Q^2)$\\
\hline
\end{tabular}
\end{center}
\caption{\label{del} Functions $\de (Q^2) = 4(\al_{\rm P}-1)$
 given by  Eq. \req{meq}. }
\end{table}

\begin{table}
{
\begin{tabular}{|lc|ccc|c|}
\hline
\multicolumn{6}{|c|}{Cross section parameters}\\
\multicolumn{6}{|c|}{$\sigma_{\rm int}(W) = C\, (W/W_0)^\delta,\; \delta= E+({F}/{1000}) \log[W/W_0]$}\\
\hline
Reaction&$Q^2$& $C$ [nb]& \multicolumn{2}{c|}{$\delta$}& $\chi^2$ \\
&[{\tiny GeV$^2$}]&&$E$&$F$&\\
$\gamma \,p \to \rho \,p $& 0 &&&&\\
\multicolumn{2}{| l|}{sect.II, wo. shrink.}&2781$\pm$ 112&0.360&0&4.28\\
\multicolumn{2}{| l|}{sect.II, w. shrink..}&5283$\pm$ 209&0.190&0&0.635\\
\multicolumn{2}{| l|}{sect.III, w. shrink.}&5362$\pm$ 212&0.181&1.3 &0.639\\ 
\multicolumn{2}{| l|}{free fit of C and $\delta$}&7351$\pm$ 1775&0.098 &0&0.110\\
\multicolumn{2}{| l|}{}&& $\pm$ 0.067&&\\
$\gamma^* \,p \to \rho \,p $& 6 &&&&\\
\multicolumn{2}{| l|}{sect.II, wo. shrink.}&11.93 $\pm$ 0.34&0.539&0&1.30\\
\multicolumn{2}{| l|}{sect.II, w. shrink.}&14.91 $\pm$ 0.37&0.486&0&0.666\\
\multicolumn{2}{| l|}{sect.III, w. shrink.}&13.26 $\pm$ 0.37&0.489&5.8&1.280\\ 
\multicolumn{2}{| l|}{free fit of C and $\delta$}&22.06 $\pm$ 9.82 & 0.393 &0&0.227\\
\multicolumn{2}{| l|}{}&& $\pm$ 0.106&&\\
$\gamma \,p \to J/\psi \,p $& 0 &&&&\\
\multicolumn{2}{| l|}{sect.II, wo. shrink.}&3.334 $\pm$ 0.045 & 0.692&0&0.845\\
\multicolumn{2}{| l|}{sect.II, w. shrink.}&3.624 $\pm$ 0.049&0.675&0&0.812\\
\multicolumn{2}{| l|}{sect.III, w. shrink.}&4.912 $\pm$ 0.067 &0.579&6.2&1.044\\ 
\multicolumn{2}{| l|}{free fit of C and $\delta$}&3.61 $\pm$ 0.35 & 0.675 &0&0.812\\
\multicolumn{2}{| l|}{}&& $\pm$ 0.020&&\\
$\gamma \,p \to \Upsilon \,p $& 0 &&&&\\
\multicolumn{2}{| l|}{sect.II, wo. shrink.}&$(8.62  \pm 1.65)10^{-4}$ & 1.04&0&0.667\\
\multicolumn{2}{| l|}{sect.III, wo. shrink.}&$(17.4 \pm 3.5) 10^{-4}$&0.862&10&0.463\\ 
\multicolumn{2}{| l|}{free fit of C and $\delta$}&$(44.6 \pm 13.7) 10^{-4}$ & 0.793 &0&0.069\\
\multicolumn{2}{| l|}{}&& $\pm$ 0.472&&\\
\hline
\end{tabular}}
\caption{\label{tab-par} Parameters of the curves displayed in Fig. \ref{cs}. The integrated elastic diffractive cross sections are  of the form
$\sigma_{\rm int}(W) = C\, (W/W_0)^\delta,\; \delta= E+({F}/{1000}) \log[W/W_0], \; W_0= 1\,  \rm{GeV}$.  The constant $C$ is always fitted to the data, the 
fixed power $\delta$ in the row ''sect.II, wo. shrink.'' is obtained from (19), in the row ''sect. II, w. shrink.'' from (25);
in the row ''sect.II, w. shrink.'' the $W$-dependent power is a parametrization of  the result obtained with the energy-dependent dipole cross section, as discussed in sect. 4, (28) ff, the shrinkage correction (24) is included.  For $\Upsilon$ production the shrinkage correction is completely negligible. For comparison we show in the row "free fit of C and $\delta$" the parameters of an unconstrained fit to the data.}
\end{table}

\subsection{high-energyBehaviour of the Dipole Model }
In this appendix we discuss the high-energybehaviour of the
dipole model of Sect. \ref{edip}. For definiteness we investigate
the simple case of elastic scattering of longitudinal virtual
photons.

The scattering amplitude is, see Eqs. (\ref{rhol}),(\ref{ampw}),  
 \beqa
\label{ap1} 
&&\cT_0 \sim i W^2 \,  \int_0^\infty d b_\perp  \,
\int_0^1  du\,
{b_\perp }\, \sigma(b_\perp,u) \times\\
&& \qquad (W/W_0)^{2 \, \be(b_\perp,u)}
 b_\perp ~  Q^2 \,u^2(1-u)^2 \, K^2_0(\ep b_\perp)   ~ . \nonumber 
\enqa 
Using as dipole cross section the simple quadratic
form $\sigma(b_\perp,u)= C b_\perp^2$ we can write Eq. \req{ap1} as
\beq \label{ap2} \cT_0 \sim  i W^2 \,  \int_0^\infty d \zeta  \,
\zeta^3\, (W/W_0)^{2 \, \be(\ze)}\ ,
 Q^2 \, K^2_0(Q \ze) ~ ,
\enq 
where we have made the phenomenologically successful
assumption that the power function $\be$ is a function of the
light-front separation $\ze= \sqrt{u(u-1)}\; b_\perp$, see 
Eq. (\ref{laga3-z}). The behaviour of the integrand $I(\zeta)$ in Eq. (\ref{ap2}) 
 near $\zeta=0$ is
$I(\zeta) \sim \ze^3 \log^2( Q\zeta)(W/W_0)^{2 \, \be(\ze)}$ and for large values $I(\zeta) \sim\ze^2 e^{-2
Q \ze}(W/W_0)^{2 \, \be(\ze)}$. We therefore investigate the integral 
\beq \label{ap3}
\cT_a = \int_0^\infty d \zeta  \, \zeta^{3-\ep}\, (W/W_0)^{2 \,
\be(\ze)}\,
 Q\,e^{ -2 Q \ze} ~. 
\enq 
We approximate  $\cT_a$ by the Gaussian integral 
\beqa
\label{ap4} \cT_a &=&  \exp[\phi(\ze_0)]\, \int_0^\infty d \zeta\,
  \exp[\frac{1}{2}(\phi''(\ze_0)(\ze-\ze_0)^2]   \nonumber \\
&\approx&\sqrt{\frac{2 \pi}{-\phi''(\ze_0)}} \;
\exp[\phi(\ze_0)], \enqa 
where 
\beq \label{ap5}
 \phi(\ze) = (3-\ep) \log \ze - Q\ze + \be(\ze) \,L; \quad L=\log\frac{W}{W_0} 
\enq 
and 
\beq \label{ap6} \phi'(\ze_0)=0  ~. \enq 
The power function
$\be(\ze)$ has a negative derivative, therefore the value of
$\ze_0$   in the limit $W\to \infty$ is driven to $\ze_0 \to 0$. 
We assume that for $\ze \to 0$ the function  $\be(\ze)$ behaves as
$\be(\ze) = \be_0 - \ga \ze^n$ . Then Eq. \req{ap6} is \beq 0 =
\frac{3-\ep}{\ze_0} - Q \ze_0 - n \ga \ze_0^{n-1} \enq and has in
the large energy limit the real root 
\beq \label{a7} \ze_0 =
\left(\frac{3-\ep}{2 n \ga L}\right)^{1/n}. \enq 
Inserting
this into Eq. (\ref{ap4})  yields for the high-energybehaviour of $\cT_a$
\beq \cT_a \sim \left(\frac{W}{W_0}\right)^{\be_0} \,
L^{-(4-\ep)/{n}} ~ .\enq

The power behaviour of $cT_a$ is independent of $\ep$
and given by the maximal value of the power  $\be(\ze)$,
and therefore also the power   of the amplitude $\cT_0$ is
given by $\be_0$. The logarithmic corrections however depend on
the specific behaviour of the function $\be(\ze)$.
%%%%%%%%%%%%%%%%%%%%%%%%%%%%%%%%%%%%%%
%%%%%%%%%%%%%%%%%%%%%%%%%%%%%%%%%%
%%%%%%%%%%%%%%%%%%%%%%%%%%%%%%%%%%%%%

%%%%%%%%%%%%%%%%%%%%%%%%%%%%%%%%%%%%
\end{document}